\documentclass[fleqn,usenatbib]{mnras}

\usepackage{newtxtext,newtxmath}
\usepackage[T1]{fontenc}

\DeclareRobustCommand{\VAN}[3]{#2}
\let\VANthebibliography\thebibliography
\def\thebibliography{\DeclareRobustCommand{\VAN}[3]{##3}\VANthebibliography}

\usepackage{graphicx}	% Including figure files
\usepackage{amsmath}	% Advanced maths commands

\usepackage{xcolor}

\newcommand{\wignerJ}[6]{\begin{pmatrix}#1 & #2 & #3\\
#4 & #5 & #6
\end{pmatrix}}

\title[ACT DR6+DESI LRGs kSZ velocity reconstruction]{The Atacama Cosmology Telescope: Large-scale velocity reconstruction with the kinematic Sunyaev--Zel'dovich effect and DESI LRGs}

\author[F. McCarthy et al.]{Fiona McCarthy$^{1,2,3}$\thanks{E-mail: fmm43@cam.ac.uk}, Nicholas Battaglia$^{4}$, Rachel Bean$^{4}$, J. Richard Bond$^{5}$, Hongbo Cai$^{6,7}$,\newauthor Erminia Calabrese$^{8}$,  William R.~Coulton$^{1,2}$, Mark J.~Devlin$^{9}$, Jo Dunkley$^{10,11}$, Simone Ferraro$^{12,13}$,\newauthor Vera Gluscevic$^{14}$,  Yilun Guan$^{15}$, J.~Colin Hill$^{16}$,
  Matthew C.~Johnson$^{17,18}$,  Aleksandra Kusiak$^{16,19,2}$,   \newauthor Alex Lagu\"e$^{9}$,
Niall MacCrann$^{1,2}$,  Mathew~S.~Madhavacheril$^{9}$, Kavilan Moodley$^{20,21}$, 
  Sigurd Naess$^{22}$, 
 \newauthor   Frank~J.~Qu$^{1,2,23}$,  
  Bernardita Ried Guachalla$^{24,23,25}$, 
 Neelima Sehgal$^{26}$, Blake D.~Sherwin$^{1,2}$,\newauthor Crist\'obal Sif\'on$^{27}$,    Kendrick M.~Smith$^{17}$, Suzanne T.~Staggs$^{10}$, Alexander van Engelen$^{28}$,\newauthor Eve M. Vavagiakis$^{29,4}$,  Edward J.~Wollack$^{30}$ 
\\
$^{1}$DAMTP, Centre for Mathematical Sciences, Wilberforce Road, Cambridge CB3 0WA, UK\\
$^{2}$Kavli Institute for Cosmology Cambridge, Madingley Road, Cambridge, CB3 0HA, UK\\
$^{3}$Center for Computational Astrophysics, Flatiron 
Institute, 162 5th Avenue, New York, NY 10010 USA\\
$^{4}$Department of Astronomy, Cornell University, Ithaca, NY, USA 14850\\
$^{5}$Canadian Institute for Theoretical Astrophysics, 60 St. George Street, University of Toronto, Toronto, ON, M5S 3H8, Canada\\
$^{6}$Department of Physics and Astronomy, University of Pittsburgh, Pittsburgh, PA, USA 15260\\
$^{7}$Department of Astronomy, School of Physics and Astronomy, Shanghai Jiao Tong University, Shanghai, 200240, China\\
$^{8}$School of Physics and Astronomy, Cardiff University, The Parade, Cardiff, Wales CF24 3AA, UK\\
$^{9}$Department of Physics and Astronomy, University of Pennsylvania, 209 South 33rd Street, Philadelphia, PA, USA 19104\\
$^{10}$Joseph Henry Laboratories of Physics, Jadwin Hall, Princeton University, Princeton, NJ, USA 08544\\
$^{11}$Department of Astrophysical Sciences, Peyton Hall, Princeton University, Princeton, NJ 08544 USA\\
$^{12}$Physics Division, Lawrence Berkeley National Laboratory, Berkeley, CA, USA\\
$^{13}$Berkeley Center for Cosmological Physics, University of California, Berkeley, CA 94720\\
$^{14}$Department of Physics and Astronomy,  University of Southern California, Los Angeles, CA, 90007, USA\\
$^{15}$Dunlap Institute for Astronomy \& Astrophysics, University of Toronto, 50 St. George St., Toronto ON M5S 3H4, Canada\\
$^{16}$Department of Physics, Columbia University, New York, NY, USA 10027\\
$^{17}$Perimeter Institute for Theoretical Physics, 31 Caroline St N, Waterloo, ON N2L 2Y5, Canada\\
$^{18}$Department of Physics and Astronomy, York University, Toronto, ON M3J 1P3, Canada\\
$^{10}$Institute of Astronomy, University of Cambridge, Madingley Road, Cambridge, CB3 0HA, UK\\
$^{20}$Astrophysics Research Centre, University of KwaZulu-Natal, Westville Campus, Durban 4041, South Africa\\
$^{21}$School of Mathematics, Statistics \& Computer Science, University of KwaZulu-Natal, Westville Campus, Durban 4041, South Africa\\
$^{22}$ Institute for theoretical astrophysics, University of Oslo,  Norway\\
$^{23}$Kavli Institute for Particle Astrophysics and Cosmology, 382 Via Pueblo Mall Stanford, CA 94305-4060, USA\\
$^{24}$Department of Physics, Stanford University, Stanford, CA, USA 94305-4085\\
$^{25}$SLAC National Accelerator Laboratory 2575 Sand Hill Road Menlo Park, California 94025, USA\\
$^{26}$Physics and Astronomy Department, Stony Brook University, Stony Brook, NY 11794, USA\\
$^{27}$Instituto de F\'isica, Pontificia Universidad Cat\'olica de Valpara\'iso, Casilla 4059, Valpara\'iso, Chile\\
$^{28}$School of Earth and Space Exploration, Arizona State University, Tempe, AZ 85287, USA\\
$^{29}$Department of Physics, Duke University, Durham, NC 27710, USA\\
$^{30}$NASA Goddard Space Flight Center, Greenbelt, MD USA 20771
}

\date{Accepted XXX. Received YYY; in original form ZZZ}

\pubyear{\the\year{}}

\begin{document}
\label{firstpage}
\pagerange{\pageref{firstpage}--\pageref{lastpage}}
\maketitle

% Abstract of the paper
\begin{abstract}

The kinematic Sunyaev--Zel'dovich (kSZ) {effect induces a non-zero density-density-temperature bispectrum, which we can use to reconstruct the large-scale velocity field from a combination of cosmic microwave background (CMB) and galaxy density measurements, in a procedure known as ``kSZ velocity reconstruction''. This method has been forecast to constrain large-scale modes with future galaxy and CMB surveys, improving their measurement beyond what is possible with the galaxy surveys alone. Such measurements will enable tighter constraints on large-scale signals such as primordial non-Gaussianity, deviations from homogeneity, and modified gravity.}  In this work, we  demonstrate a statistically significant measurement of kSZ velocity reconstruction for the first time, by applying quadratic estimators to the combination of the ACT DR6 CMB+kSZ map and the DESI LRG galaxies (with photometric redshifts) in order to reconstruct the velocity field. {We do so using a formalism appropriate for the 2-dimensional projected galaxy fields that we use, which naturally incorporates the curved-sky effects important on the largest scales.} We find evidence for the signal  by cross-correlating with an external estimate of the velocity field from the spectroscopic BOSS survey  and rejecting the null (no-kSZ) hypothesis at $3.8\sigma$. Our work presents a first step towards the use of this observable for cosmological analyses.

\end{abstract}

% Select between one and six entries from the list of approved keywords.
% Don't make up new ones.
%\begin{keywords}%
%keyword1 -- keyword2 -- keyword3
%\end{keywords}

%%%%%%%%%%%%%%%%%%%%%%%%%%%%%%%%%%%%%%%%%%%%%%%%%%

%%%%%%%%%%%%%%%%% BODY OF PAPER %%%%%%%%%%%%%%%%%%

\section{Introduction}

Small-scale measurements of the cosmic microwave background (CMB) anisotropies are dominated by the CMB secondary anisotropies---chiefly the Sunyaev--Zel'dovich (SZ) effects~\citep{1969Ap&SS...4..301Z,1970Ap&SS...7....3S,1980MNRAS.190..413S}  and CMB lensing~\citep{2006PhR...429....1L,2007PhRvD..76d3510S}. The former arise from the electromagnetic interaction (scattering) of CMB photons with electrons along the line of sight, and the latter from their gravitational interactions with matter. As we measure the CMB at higher resolution with ground-based experiments like the Atacama Cosmology Telescope (ACT)~\citep{2007ApOpt..46.3444F,2016JLTP..184..772H}, the South Pole Telescope (SPT)~\citep{2011PASP..123..568C,2014SPIE.9153E..1PB}, Simons Observatory (SO)~\citep{2019JCAP...02..056A}, and CMB-S4~\citep{2016arXiv161002743A}, we will continue to make extremely high signal-to-noise measurements of these signals, which contain different and complementary information to the primary anisotropies that dominated the space-based instrument era.

The SZ interaction has several different 
 effects on CMB maps, depending on the properties of the electron gas that scatters the CMB. The {thermal} SZ (tSZ) effect is sourced when the electrons have relatively high pressure and upscatter the CMB photon; this induces a well-understood spectral distortion in the CMB~\citep{1970Ap&SS...7....3S}, and thus can be isolated with its frequency dependence (see, e.g.~\citealt{2014A&A...571A..21P,2014JCAP...02..030H,2016A&A...594A..22P,2020PhRvD.102b3534M,2022ApJS..258...36B,2022MNRAS.509..300T,2023MNRAS.526.5682C,2024PhRvD.109b3528M}). The tSZ effect is a well-established probe of the thermodynamics of the highly massive, late-Universe objects (halos hosting galaxy clusters) that source the signal (see, e.g.~\citealt{1978Natur.275...40B,1978MNRAS.185..245B,1981MNRAS.197..571B,2019A&A...622A.136H,2021PhRvD.103f3514A,2021PhRvD.104d3503V,2022PhRvD.105l3525G,2022PhRvD.105l3526P}), and of the distributions of these clusters (see, e.g.~\citealt{2016A&A...594A..24P,2019MNRAS.489..401Z,2022ApJ...934..129S,2024arXiv240102075B}). See ~\cite{2002ARA&A..40..643C,2019SSRv..215...17M} for reviews.

The {kinematic} SZ effect, in contrast, is sourced by the scattering of CMB photons off a gas of electrons that has a non-zero bulk velocity with respect to the CMB; the signal can be thought of as the Doppler shifting of the CMB. In this case, no spectral distortion is imparted to the photons, and so this signal cannot be separated from the primary CMB using multifrequency measurements. The primary anisotropies  are  dominant on large-to-intermediate scales, making measurements of the  kSZ signal difficult without  high resolution data. Detections have been made with increasing significance through cross-correlations (see, e.g.~\citealt{2012PhRvL.109d1101H,2013A&A...557A..52P,2016PhRvD..93h2002S,2016MNRAS.461.3172S,2016PhRvL.117e1301H,2021PhRvD.103f3513S,2021PhRvD.104d3518K,2021A&A...645A.112T,2021PhRvD.104d3502C,2024arXiv240707152H}).

\begin{figure*}
\includegraphics[width=0.49\textwidth]{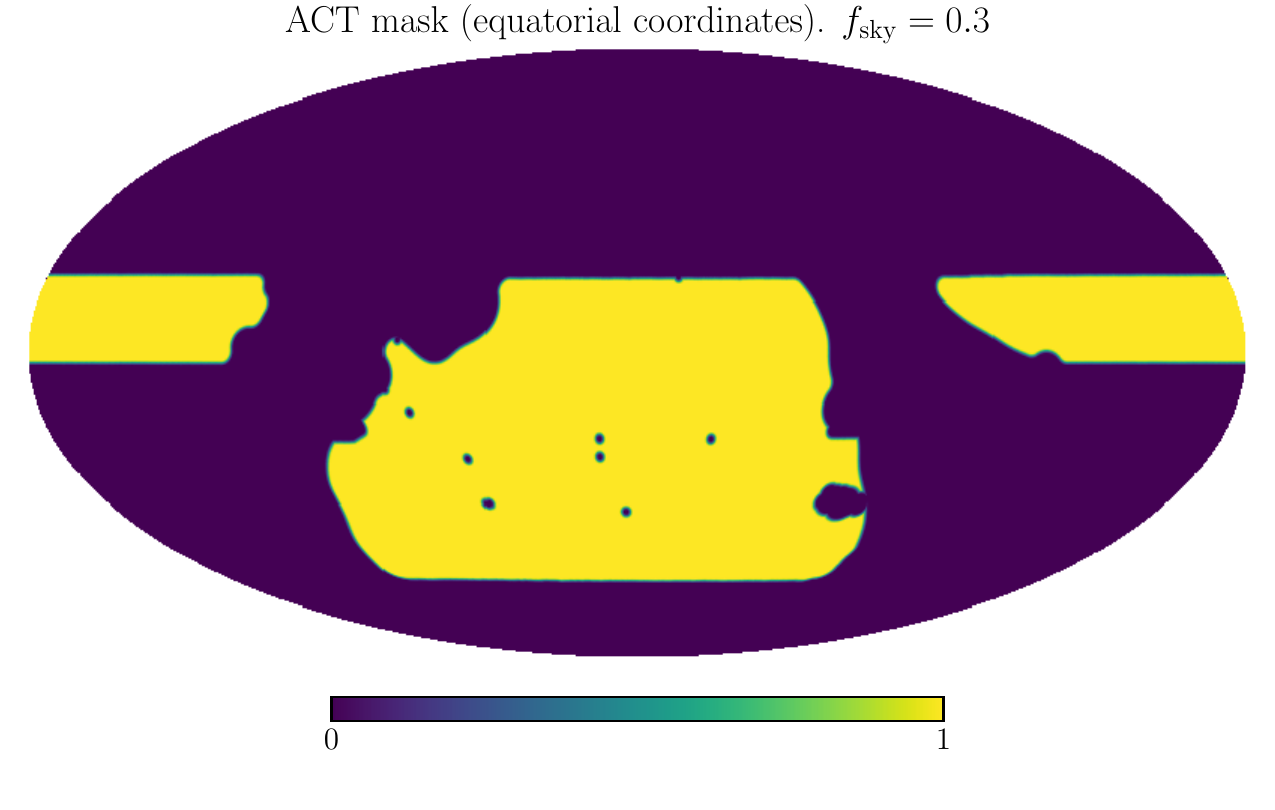}
\includegraphics[width=0.49\textwidth]{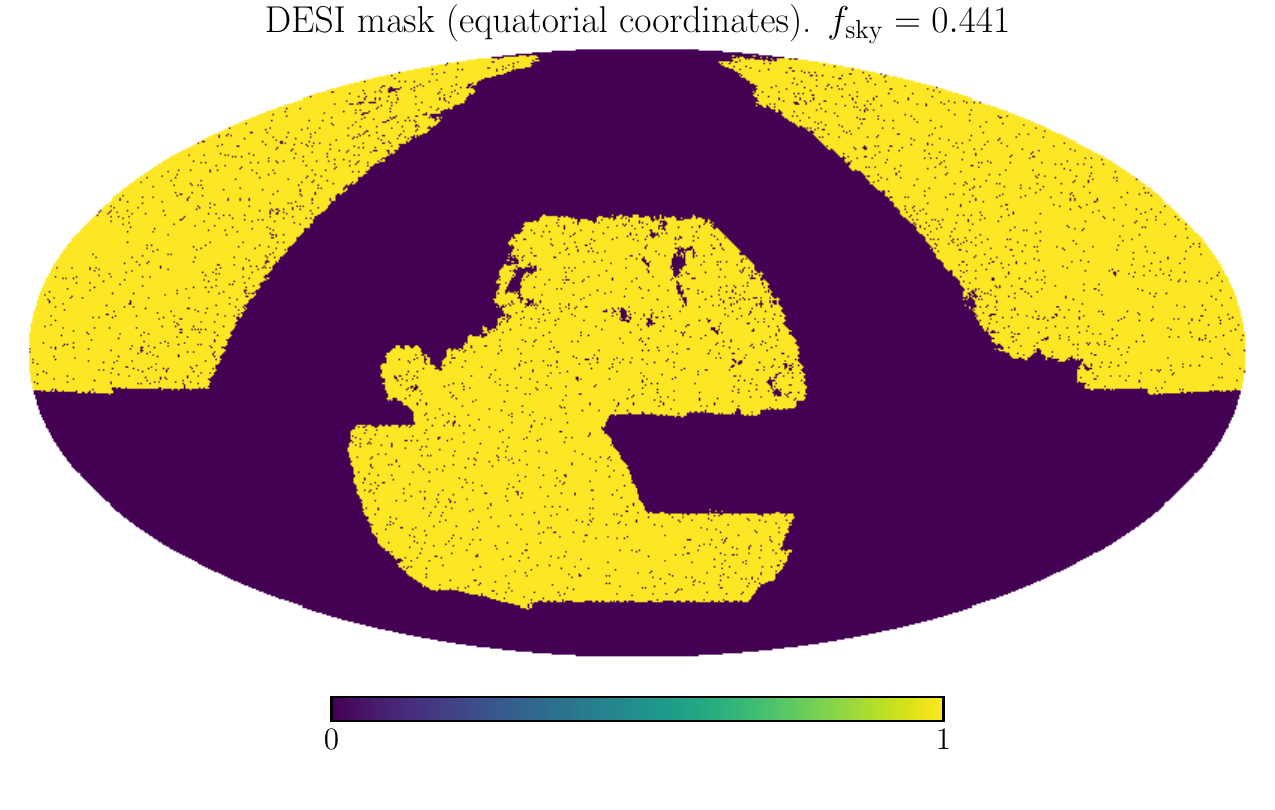}
\includegraphics[width=0.49\textwidth]{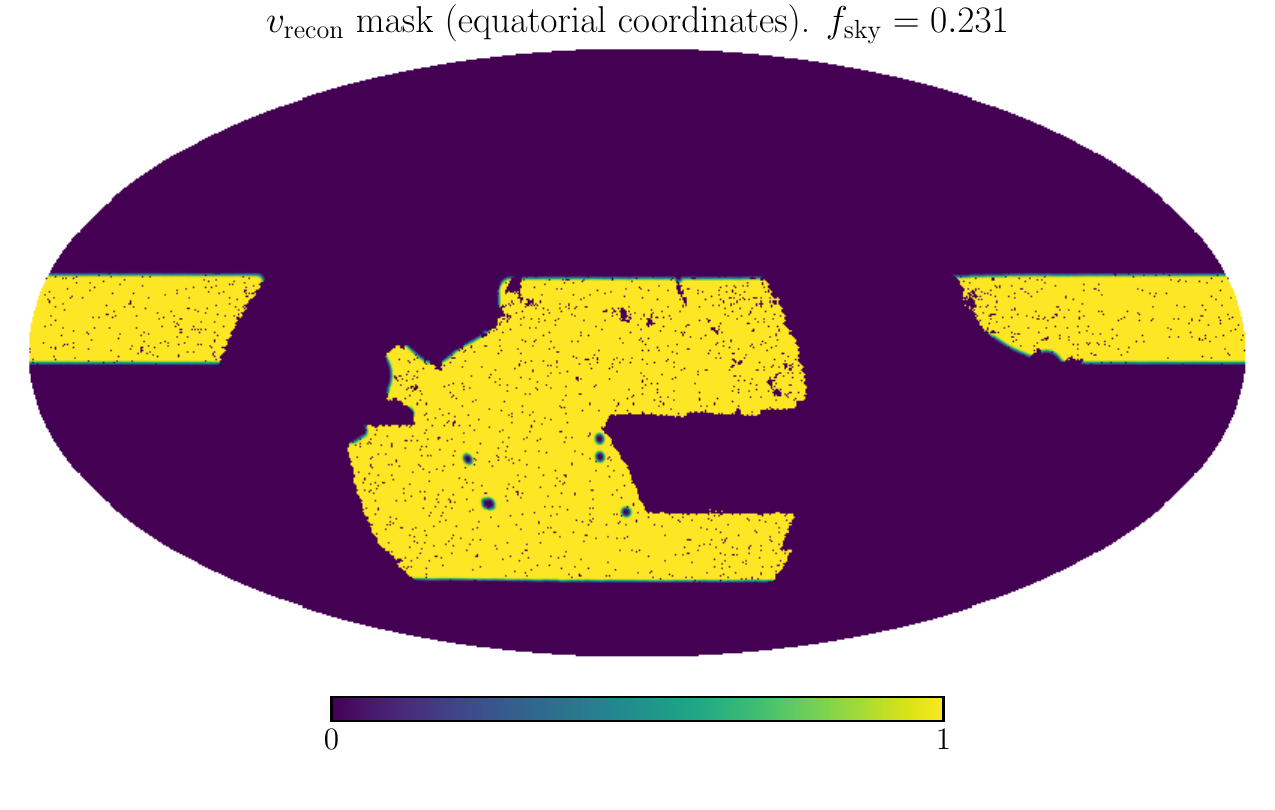}
\includegraphics[width=0.49\textwidth]{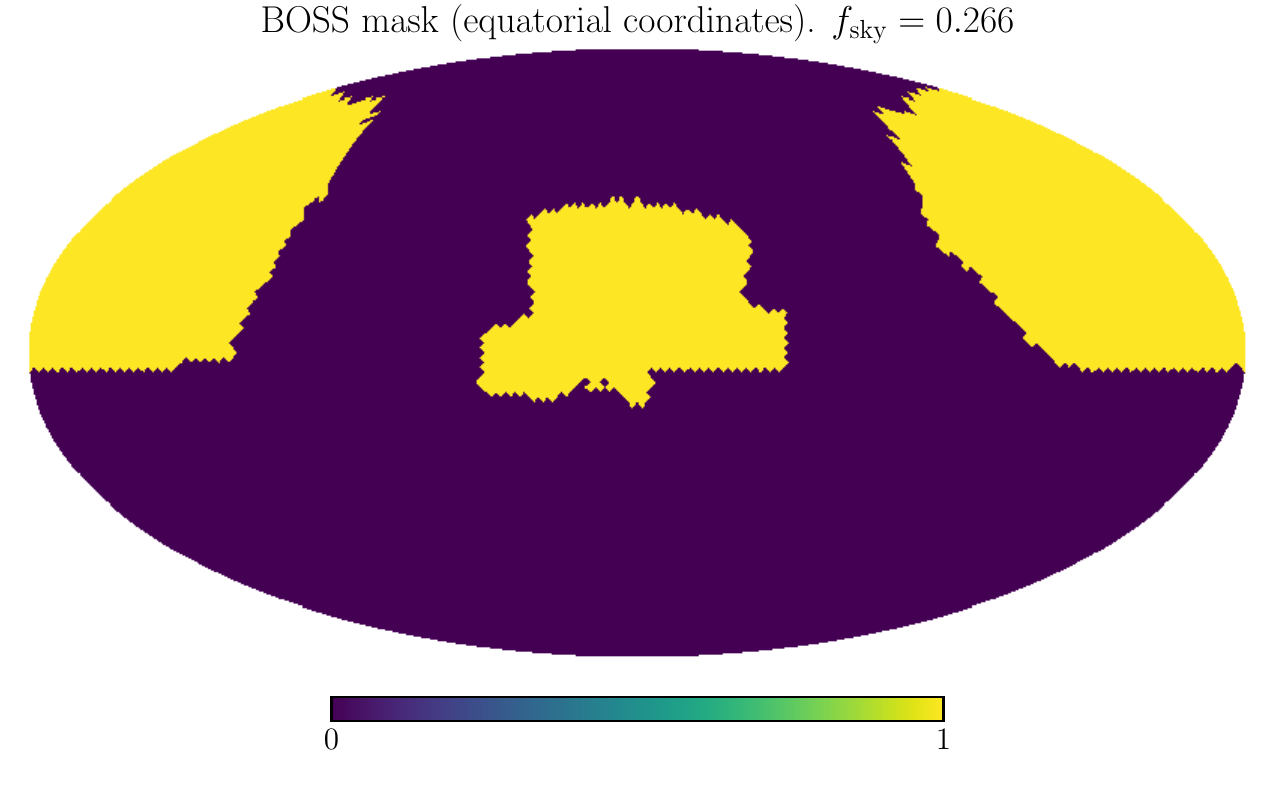}

\caption{The ACT mask (\textit{top left}) and the DESI mask (\textit{top right}) applied respectively to the kSZ map and each galaxy overdensity map for the velocity reconstruction. The resulting reconstructed velocity has their combined mask applied (\textit{bottom left}) with an overlap area of 23.1\% of the sky. The mask applied to the BOSS template is also shown (\textit{bottom right}), with a sky area of 26.6\%. The overlap area of the kSZ reconstruction mask and the BOSS mask is $f_{\mathrm{sky}}=11\%$. }\label{fig:masks}

\end{figure*}

The kSZ signal depends on two properties of the electron gas that scatters the CMB: its density and radial velocity (with respect to the CMB). For most kSZ observables, the relevant quantities that source the anisotropies are the small-scale fluctuations in the electron density, and the \textit{large-scale} fluctuations in the velocity field.  By using an external measurement of the electron velocity (e.g. by inferring velocity from the large-scale density field), the velocity-electron density degeneracy can be broken, and the kSZ signal can become a probe of the electron density of the Universe. Similarly, combination with an external measurement of the electron density allows the degeneracy to be broken to allow for direct measurements of the large-scale velocity. Such a measurement has been recognized to have the power to probe the properties of the Universe on extremely large scales, through a process known as ``kSZ tomography'', or ``kSZ velocity reconstruction''(see, e.g.~\citealt{Ho:2009iw,Munshi:2015anr,Zhang_2010,Shao_2011,2017JCAP...02..040T,2018PhRvD..98l3501D,2018arXiv181013423S}). This property of the signal has been forecast to be sensitive to local primordial non-Gaussianities, large-scale homogeneity and anisotropy, and modified gravity~\citep{2015JCAP...06..046Z,2019PhRvD.100h3508M,2019JCAP...10..024C,2019PhRvD.100h3522P,2020PhRvD.101l3508C,2023PhRvD.107d3504K,2023PhRvD.108l3528C}. 

kSZ tomography has been demonstrated with simulations~\citep{2018PhRvD..98f3502C,2022JCAP...09..028G,2023JCAP...02..051C} and applied to \textit{Planck} CMB data~\citep{2024arXiv240500809B}, with the {unWISE} galaxy sample~\citep{2020JCAP...05..047K,2021JCAP...12..028K} used as the tracer of electron density. This combination led to a signal consistent with zero at $\sim1\sigma$. In this work, we use higher-resolution CMB data from ACT DR6~\citep{2024PhRvD.109f3530C}, along with the Dark Energy Spectroscopic Instrument (DESI) luminous red galaxy (LRG)  sample~\citep{2023JCAP...11..097Z}, to perform kSZ velocity reconstruction. While there are more objects, with a higher number density, in the {unWISE} sample, their very wide redshift distributions (and large redshift uncertainties) significantly dilute the velocity signal. The DESI LRGs, in contrast, have the advantage of having per-object photo-$z$ estimates, which can be used to bin the velocity tomographically and reduce the line-of-sight cancellation. Additionally, due to the large overlap with the BOSS spectroscopic survey, we can benefit from making a detection using a cross-correlation with an external measurement of the velocity field, reconstructed from the continuity equation applied to the BOSS galaxies. We find $\sim3.8\sigma$ evidence of the cross-correlation between our kSZ-estimated velocity and this template.

This is a first step towards the application of a precision analysis pipeline to such signals. In particular, the signal we measure is sensitive to local primordial non-Gaussianity (through the continuity equation velocity estimate) and can be directly used to constrain $f_{\mathrm{NL}}^{\mathrm{loc}}$. In the broader landscape of upcoming high-resolution, low-noise CMB surveys such as SO and CMB-S4 and upcoming high-number-density galaxy surveys, including photometric surveys such as the Vera Rubin Observatory's LSST~\citep{2009arXiv0912.0201L}, Euclid~\citep{2013LRR....16....6A}, and  SPHEREX~\citep{2014arXiv1412.4872D}, as well as spectroscopic surveys like DESI~\citep{2016arXiv161100036D}, kSZ tomography will turn into a high-signal-to-noise cosmological probe.

The $\left<Tgg\right>$ statistic (where the first $T$ is the CMB temperature, one $g$ is the small-scale galaxy density, and the other $g$ is the large-scale velocity field that is inferred from a galaxy survey) is the same statistic that is probed in kSZ velocity-weighted \textit{stacking}, e.g. in~\cite{2021PhRvD.103f3513S,2024arXiv240707152H}, where detections of the signal were made at $6.5\sigma$ and $13\sigma$ respectively, using the data combinations of ACT DR5+BOSS; ACT DR6+DESI LRGs respectively. However, the stacking measurement is more appropriate for constraining the baryon distribution, while our kSZ velocity reconstruction analysis method is more appropriate for constraining large-scale cosmology~\citep{2018arXiv181013423S}. These detections show that with various data combinations, detections of the large-scale velocity field with higher significance will be possible imminently.

For all cosmological calculations throughout, we use the \textit{Planck} 2018 cosmology~\citep{2020A&A...641A...6P}, where $H_0\equiv 100h=67.66\, \mathrm{km/s/Mpc}$, $\Omega_b h^2=0.02242$, $\Omega_c h^2=0.11933$,  $\ln(10^{10}A_s)=3.047$, and $n_s=0.9665$, with $H_0$ the Hubble parameter today, $\Omega_b h^2$ the physical baryon density, $\Omega_c h^2$ the physical cold dark matter density, and $A_s$ and $n_s$ the amplitude and spectral index of scalar fluctuations respectively (at a pivot scale of $0.05\, \mathrm{Mpc}^{-1}$).

\section{Data}

We use three datasets: a CMB blackbody map from ACT DR6+\textit{Planck}~\citep{2024PhRvD.109f3530C} and a photometric galaxy sample from the DESI Legacy Survey~\citep{2019AJ....157..168D} for the kSZ velocity reconstruction; and a spectroscopic galaxy sample from SDSS for the large-scale velocity template. We describe these below.

\subsection{kSZ dataset: ACT DR6}
The CMB map we use is the ACT DR6+\textit{Planck}  CMB temperature map, which is a needlet internal linear combination (NILC)~\citep{2009A&A...493..835D} estimation of the blackbody signal in the multifrequency ACT data, described in~\cite{2024PhRvD.109f3530C}. {NILC allows for the creation of a minimum-variance map from separate multifrequency maps which preserves the frequency dependence of a signal of interest (in this case, the blackbody primary CMB+kSZ signal) and allows for inter-frequency cleaning of other signals (such as the tSZ signal and other foregrounds).} The map is created from a combination of ACT DR4 and DR6 data and \textit{Planck} NPIPE data~\citep{2020A&A...643A..42P}. It covers 30\% of the sky and is convolved with a 1.6 arcminute {Gaussian} beam. The area of sky covered is shown in Figure~\ref{fig:masks}. We reproject the map from the Plate-Carr\'{e}e (CAR) coordinate system to a HEALPix coordinate system, with an $N_{\mathrm{side}}$ of 2048 (which has pixel size $\sim1.7^\prime$), by first projecting to spherical harmonics with $\ell_{\mathrm{max}}=3\times2048$ and then to HEALPix. We note that the downgrading to $N_{\mathrm{side}}=2048$, which is lower resolution than the native $\ell_{\max}\sim21000$ resolution of the ACT DR6 map, limits us to a maximum multipole $\ell_{\mathrm{max}}=3\times2048=6144$. While the kSZ signal is stronger on smaller scales, due to residual foreground power the improvement at high-$\ell$ is limited, and with this cut-off we capture $\sim95\%$ of the signal-to-noise.

\subsection{Photometric galaxy dataset}
The galaxy dataset we use for kSZ velocity reconstruction is the DESI LRG sample with photometric redshifts described in~\cite{2023JCAP...11..097Z}. These objects were selected from the imaging data from the DESI Legacy Survey DR9~\citep{2019AJ....157..168D,2023AJ....165...58Z}.

\begin{figure}
\includegraphics[width=\columnwidth]{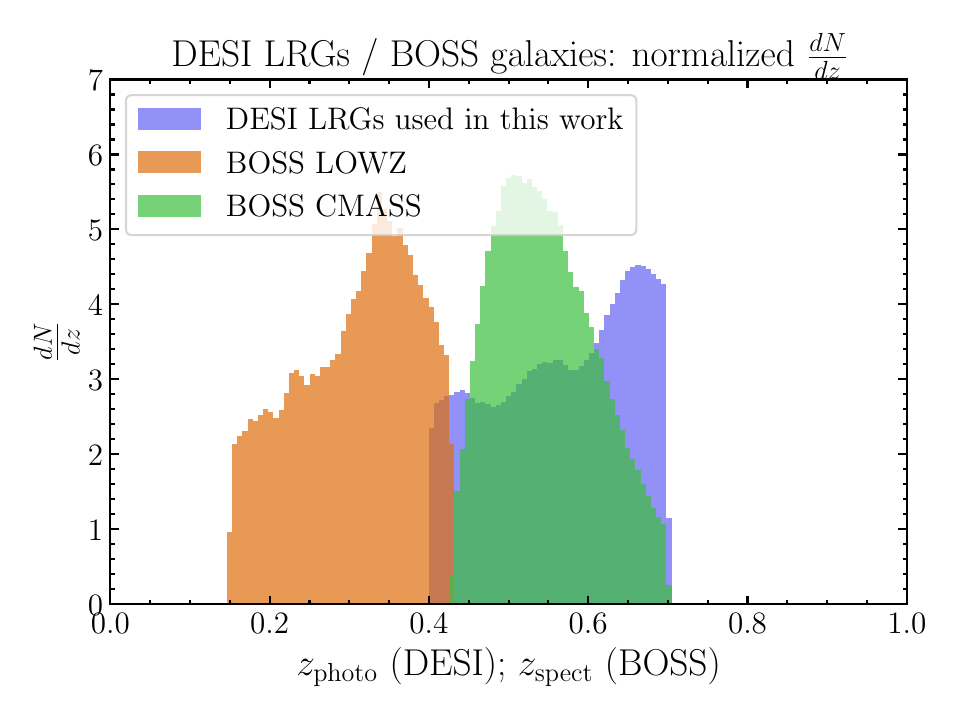}
\caption{The ${z^{\mathrm{photo}}}$ distribution  for the DESI LRGs that we use in this work  is shown (\textit{blue}) along with the distribution of \textit{spectroscopic} redshifts of the BOSS LOWZ (\textit{orange}) and CMASS (\textit{green}) galaxies.  In every case the normalization is such that the distribution integrated  over the redshift range is 1.
}\label{fig:dndz_all}
\end{figure}

In~\cite{2023JCAP...11..097Z}, the subset of these galaxies with photo-$z$ in the range $0.4<z_{\mathrm{photo}}<1$  was split into four tomographic redshift bins with spectroscopically-confirmed redshift distributions, for use in measurements that benefit from tomography and require accurate redshift distributions (such as for the ACT DR6 CMB lensing cross correlation analyses of~\citealt{2024arXiv240704606K,2024arXiv240704607S}). We similarly impose a $z_{\mathrm{photo}}>0.4$ cut in order to avoid modelling the redshift distribution in the tails of the sample. We also require that the objects overlap in redshift with the BOSS galaxies, to perform the cross-correlation with BOSS galaxies, and so only use objects with $z_{\mathrm{photo}}<0.7$. There are 9,638,964 such objects (out of 33,735,219 in the entire catalog). The distribution of their photometric redshifts is shown in Figure~\ref{fig:dndz_all}.

We rebin these galaxies into four photometric redshift bins such that there are an equal number (2,409,741) of galaxies in each bin. Thus the photometric redshift boundaries are $[0.4,0.49,0.57,0.64,0.7]$. To model the \textit{true} redshift distributions, we convolve these step (top-hat) functions with Gaussians with width $0.027\times(1+z)$, which quantifies the photometric redshift uncertainty; these true redshift distribution estimates are also shown in Figure~\ref{fig:dndz_4bins}.  Thus, the $\frac{dN}{dz}$ are approximate; in order to mitigate mismodelling of this redshift distribution, we will not analyse any inter-bin measurements, as these are more sensitive to the poorly-modelled tails of the redshift distributions.

Note that our photometric bins are different from the four photometric bins defined in~\cite{2023JCAP...11..097Z} with spectroscopic calibration, as we require bins with finer redshift resolution in order to maximize the signal-to-noise. We indicate in  Figure~\ref{fig:dndz_4bins} a vertical grid in $z$ with 100 [comoving] Mpc separation. This is the approximate velocity coherence length. As our bins are slightly wider than this, we still have some dilution of the velocity signal.

We project the galaxy bins onto the sphere and convert the projected galaxy  maps $g^i(\hat n)$ into galaxy overdensity maps $\delta^{g^i}$ according to
\begin{equation}
\delta^{g^i}(\hat n) = \frac{g^i(\hat n) - \bar n^i}{\bar n^i}
\end{equation}
where $\bar n^i$ is the mean galaxy density in bin $i$, which is measured directly from the projected maps by dividing the number of objects by the sky area.

\subsection{Spectroscopic galaxy dataset}\label{sec:boss_description}

\subsubsection{3-D velocities from SDSS-BOSS galaxies}
\begin{figure}
\includegraphics[width=\columnwidth]{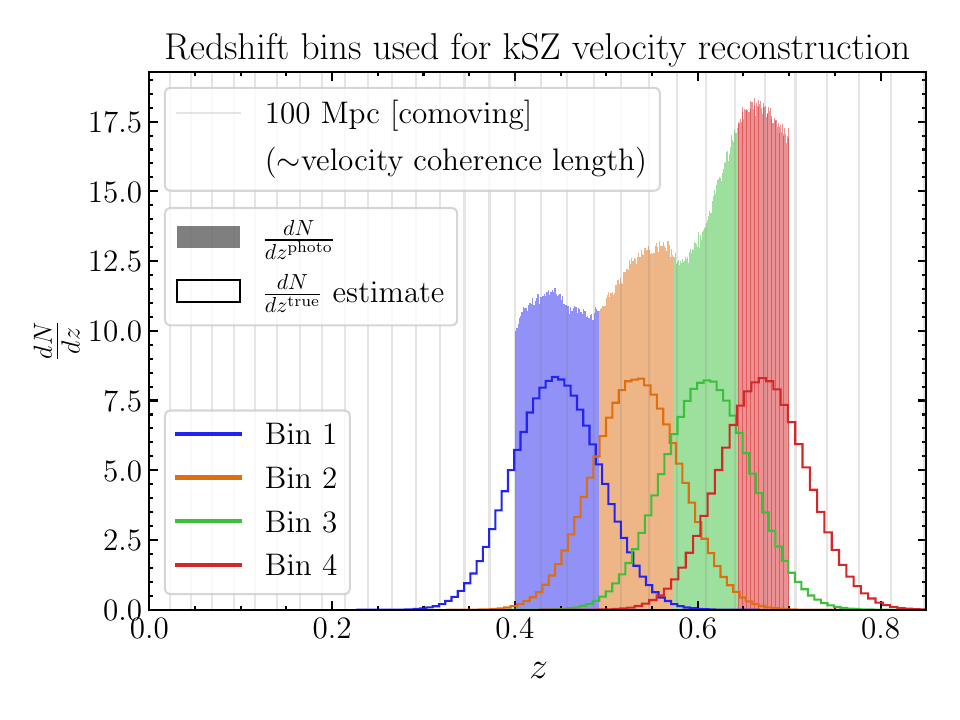}
\caption{The four DESI LRG redshift distributions used for kSZ velocity reconstruction binned according to photometric redshift (\textit{shaded}). The true $\frac{dN}{dz}$ for these samples is modeled by convolving photometric $\frac{dN}{dz^{\mathrm{photo}}}$ with a Gaussian with width $0.027\times(1+z)$ (\textit{solid lines}). All distributions are normalized such that they integrate to 1. Note that the entire $\frac{dN}{dz^{\mathrm{photo}}}$ distribution is identical (up to normalization) to the blue distribution in Figure~\ref{fig:dndz_all}.}\label{fig:dndz_4bins}
\end{figure}
\begin{figure*}
\includegraphics[width=0.49\textwidth]{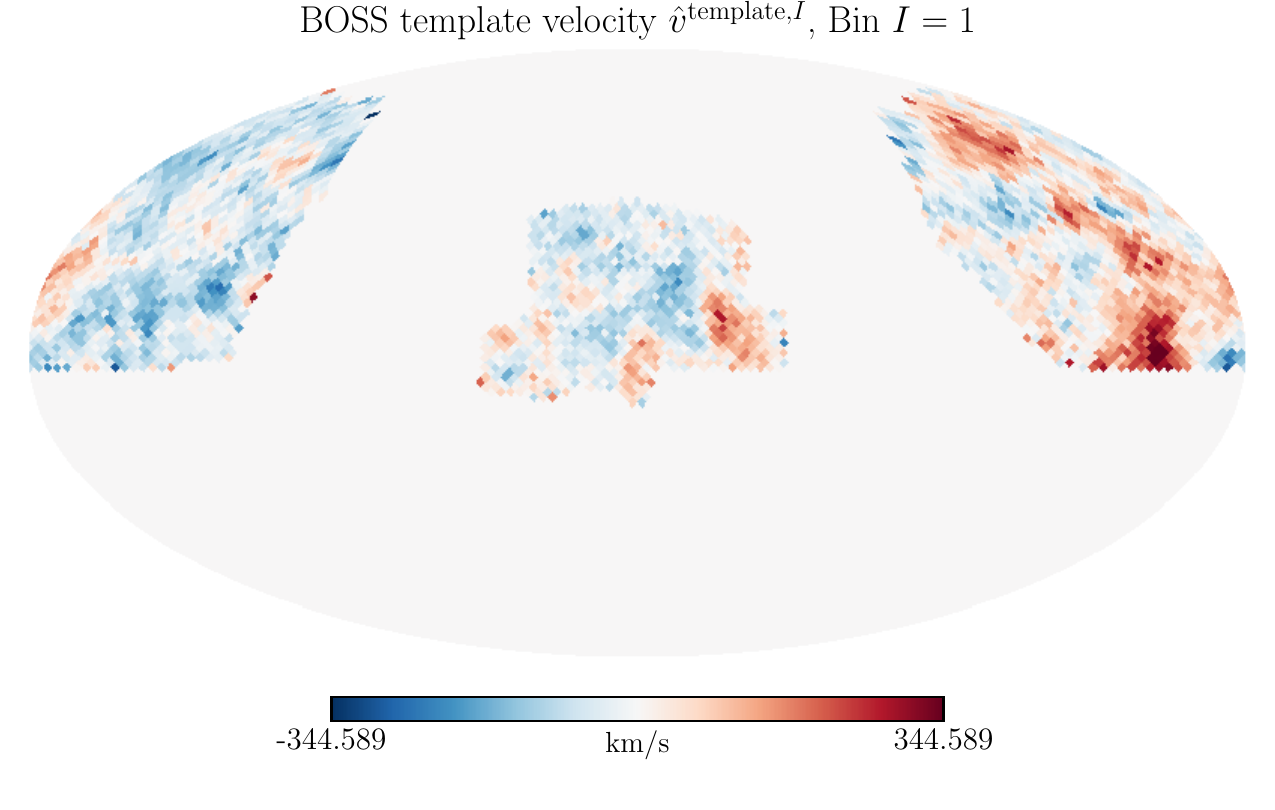}
\includegraphics[width=0.49\textwidth]{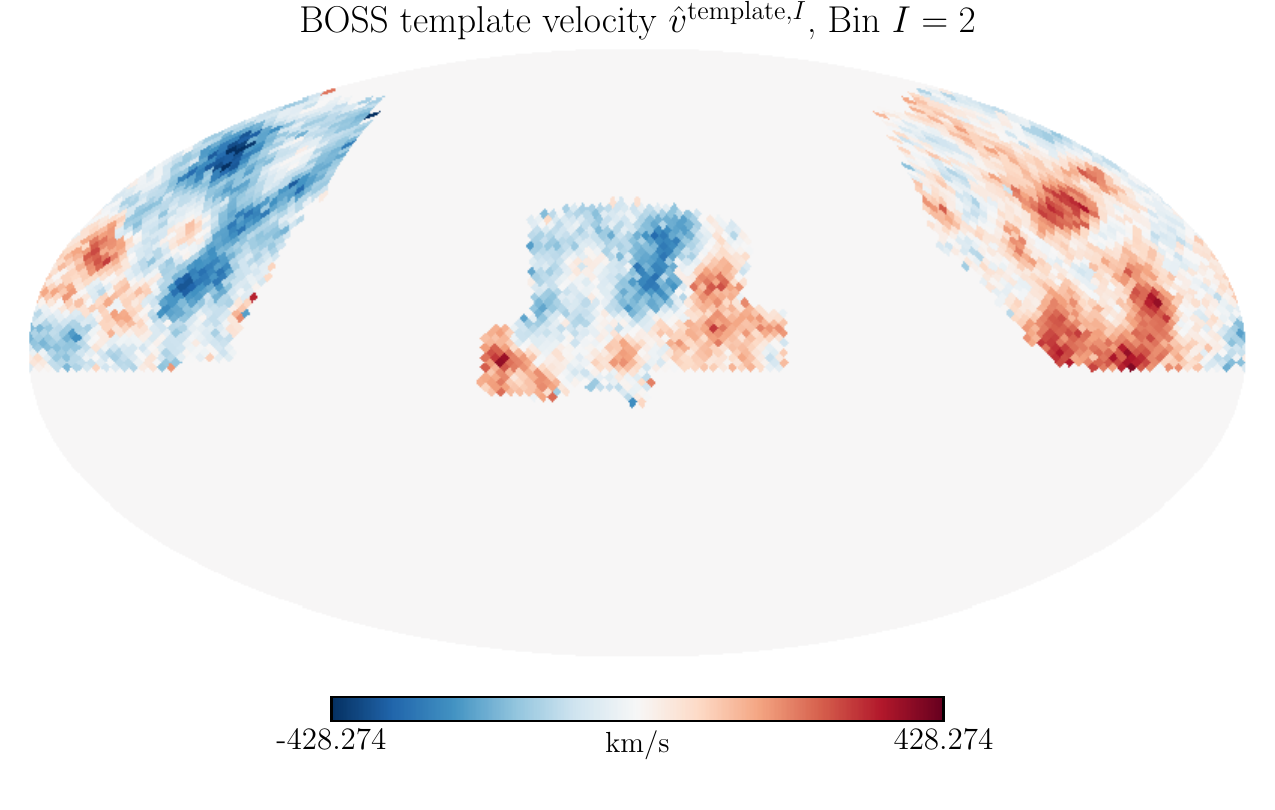}
\includegraphics[width=0.49\textwidth]{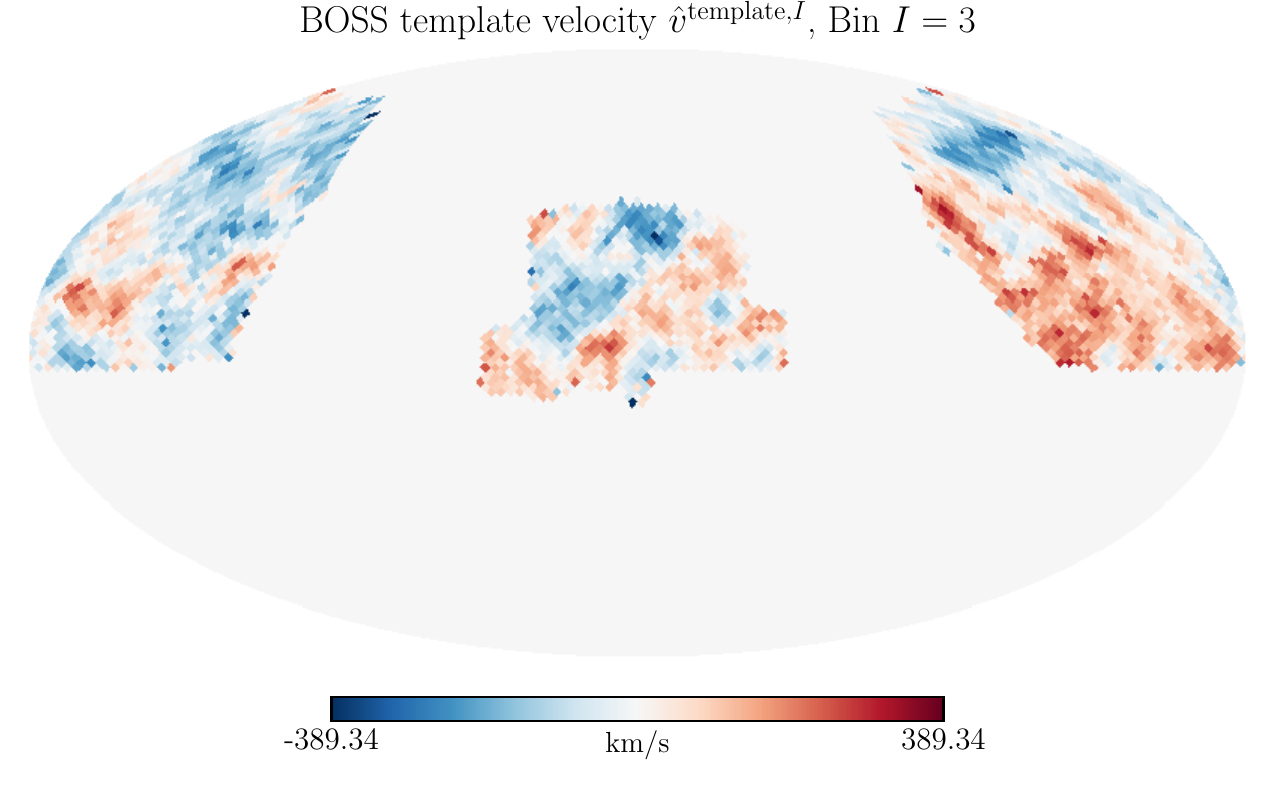}
\includegraphics[width=0.49\textwidth]{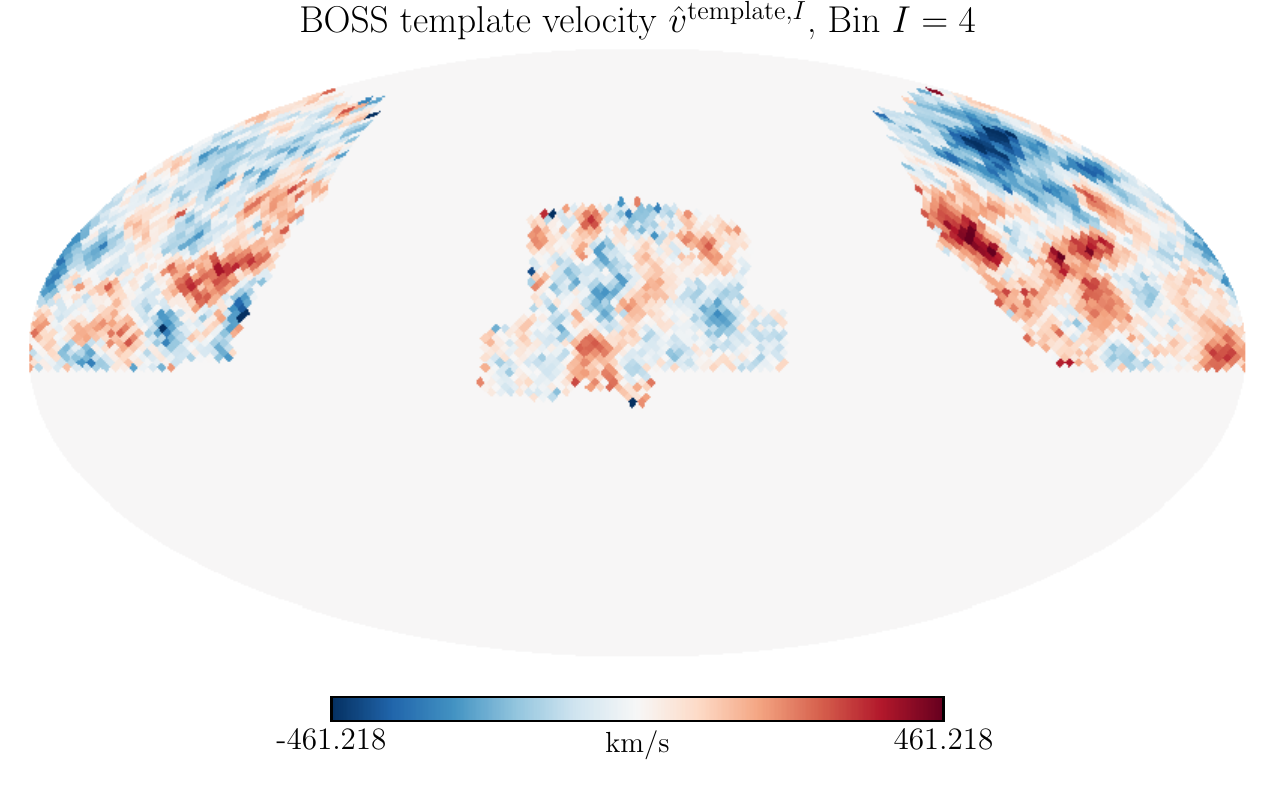}

\caption{The radial velocity template maps created from the continuity equation applied to BOSS data, as described in Section~\ref{sec:boss_description}.}\label{fig:template_maps}
\end{figure*}
For the velocity template we use in the cross-correlation, we use velocities reconstructed from the SDSS DR10 BOSS  galaxy sample~\citep{2014ApJS..211...17A}. We use the filtered dataset as described in~\cite{2021PhRvD.103f3513S} which has had velocity reconstruction 
performed on it using the \textit{continuity equation}, which infers velocity from overdensity according to
\begin{equation}
 \vec \nabla \cdot \vec v^{\mathrm{cont}} + \frac{f}{b} \vec \nabla \cdot \left[  \left( \vec v^{\mathrm{cont}} \cdot \hat n \right)\hat n \right] = - a H f \frac{\delta^g}{b},
\end{equation}
where $\vec v^{\mathrm{cont}}$ is the three-dimensional velocity that appears in the continuity equation; $f\equiv\frac{d\ln \delta}{d\ln a}$ (with $\delta$ the dark matter overdensity and $a$ the scale factor) is the growth rate; $\hat n$ is a unit vector centered on the origin describing direction; $H$ is the Hubble parameter; $\delta^g$ is the galaxy overdensity; and $b$ is the linear galaxy bias.   We use galaxies both from the CMASS and LOWZ samples.

\subsubsection{Creation of 2-D velocity templates}

We take the three-dimensional velocity field evaluated at the positions of the  galaxies, and project it into the 2-dimensional radial velocity field on HEALPix maps of $N_{\mathrm{side}}=32$ (pixel size $\sim1.8^\circ$). We refer to this as $v_r ^{\rm{BOSS,\mathrm{cont}}}$. In each pixel, we take the mean value of the radial velocity  of the galaxies in that pixel reweighted by the relative redshift distributions between the BOSS samples and the four DESI LRG bins that we will use to perform kSZ velocity reconstruction. Explicitly, we have four templates (labelled by $I$):
\begin{equation}
\hat v_r^{\mathrm{template}, I} = \frac{\int v_r ^{\rm{BOSS,\mathrm{cont}}} \frac{dN^{I}}{dz}dz}{\int \frac{dN^{\mathrm{BOSS}}}{dz}dz}.
\end{equation}
where $\frac{dN^I}{dz}$ is the normalized redshift distribution of the galaxies in bin $I$ (over which we want to estimate the velocities) and $ \frac{dN^{\mathrm{BOSS}}}{dz}$ is the normalized redshift distribution of the BOSS galaxies.

\subsubsection{Velocity template normalization}

The BOSS velocities we use are Wiener-filtered. Normally, their cross-correlation with our kSZ reconstruction should be normalized by Monte-Carlo methods.  However, in our case, we apply a correction by comparing the power spectra of the measurements to those expected from theory.
 In particular, we show the template velocity auto power spectra for each bin  in Figure~\ref{fig:autospectra_template} of Appendix~\ref{app:velocity_normalization}. We can then rescale the velocities by an amplitude $A_v^I$ such that they match the theory in each redshift bin. By comparing the measurements with theory, we find that the velocities should be rescaled by the values listed in Table~\ref{tab:velocity_rescaling_factors}.

 \begin{table}
 \centering
\begin{tabular}{|c|c|}\hline
Bin $I$& $A_v^I$\\\hline
1 &0.86\\
2 &0.70\\
3 &0.77\\ 
4 &0.65\\\hline
\end{tabular}
\caption{The normalizations $A_v^I$ which we apply to the template velocities to account for the Wiener-filtering (we rescale the templates by $1/A_v^I$).}\label{tab:velocity_rescaling_factors}
 \end{table}

We show  the four radial velocity templates in Figure~\ref{fig:template_maps}.

\section{Reconstruction formalism and pipeline}

To perform kSZ velocity reconstruction, we use an adapted version of the public code \texttt{ReCCO}~\citep{2023JCAP...02..051C}.~\footnote{\url{https://github.com/jcayuso/ReCCO}} A very general version of the velocity estimator we use is described in detail in that reference. A similar version has previously been applied to the combination of \textit{Planck} kSZ + unWISE galaxy data in~\cite{2024arXiv240500809B}.  We describe the estimator in this section.

We work in the 2-dimensional ``lightcone'' formalism appropriate for kSZ velocity reconstruction with photometric galaxies for which we do not have 3-dimensional information (or for which we can gain quasi-3-dimensional information by using tomographic redshift bins). This is in contrast to the 3-dimensional formalism which is more appropriate to spectroscopic surveys for which the full three-dimensional information is known (see, e.g.,~\citealt{2018arXiv181013423S}).

The curved-sky harmonic space estimator for kSZ velocity reconstruction with galaxies was introduced in~\cite{2018PhRvD..98l3501D} and uses formalism that was developed for CMB lensing reconstruction in~\cite{2003PhRvD..67h3002O}. In both the CMB lensing and kSZ velocity reconstructions, the estimators search for a distinctive statistically anisotropic signature induced in the statistically isotropic CMB field through the modulation of small-scale statistics by a large-scale mode.  In the case of CMB lensing, the small scale CMB power spectrum is modulated by the large-scale {lensing convergence and shear}. For the kSZ velocity reconstruction, the small-scale galaxy-electron power spectrum is modulated by the large-scale radial velocity field.

In practice, we use a real-space version of the harmonic space estimator, which first appeared in~\cite{2018PhRvD..98f3502C} (and was generalized in~\cite{2023JCAP...02..051C}), as it is computationally more efficient and more practical to implement.

\begin{figure*}
\includegraphics[width=0.49\textwidth]{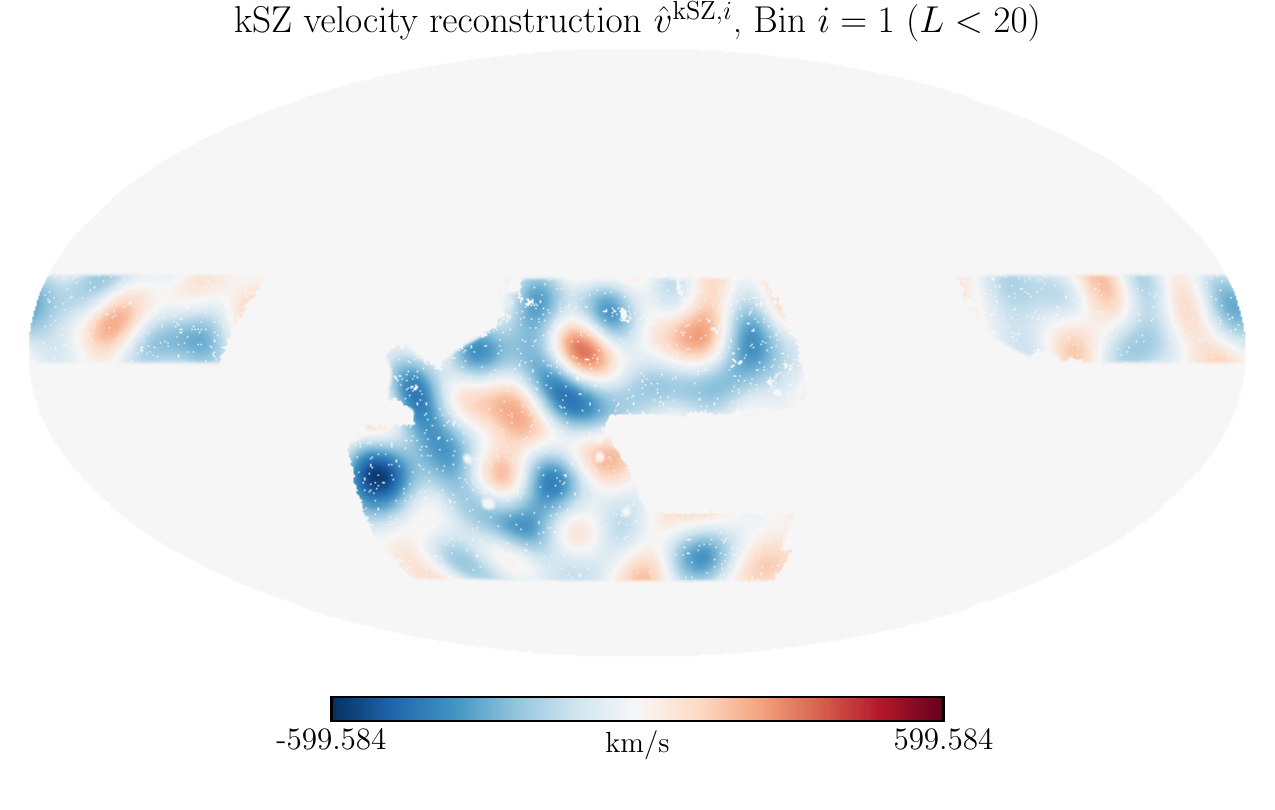}
\includegraphics[width=0.49\textwidth]{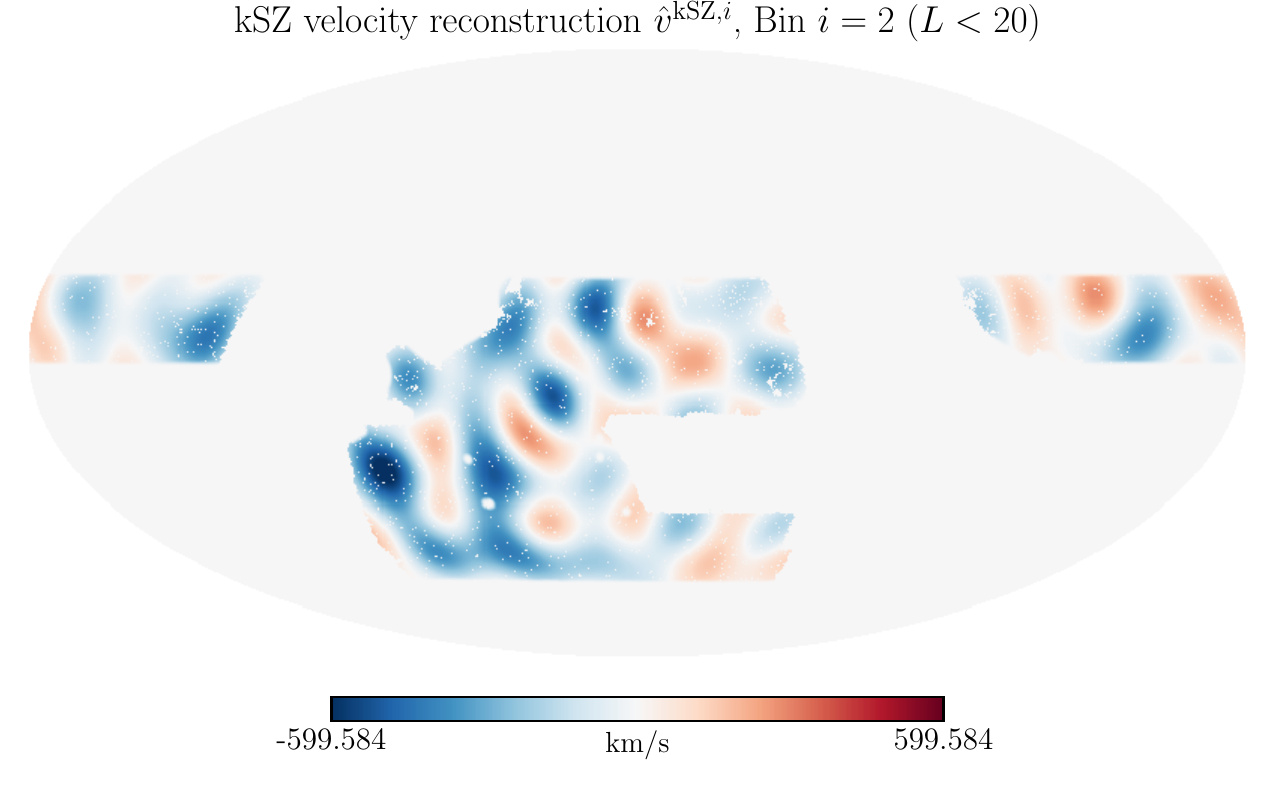}
\includegraphics[width=0.49\textwidth]{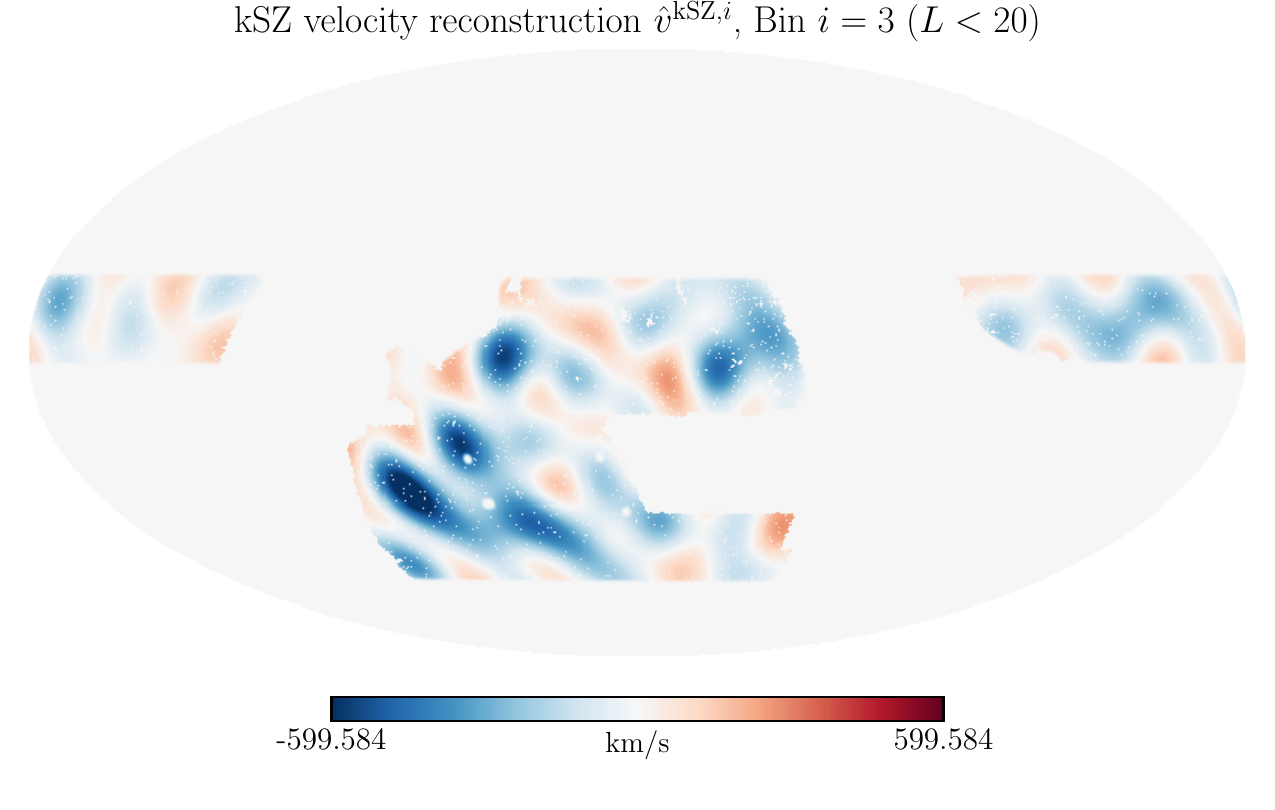}
\includegraphics[width=0.49\textwidth]{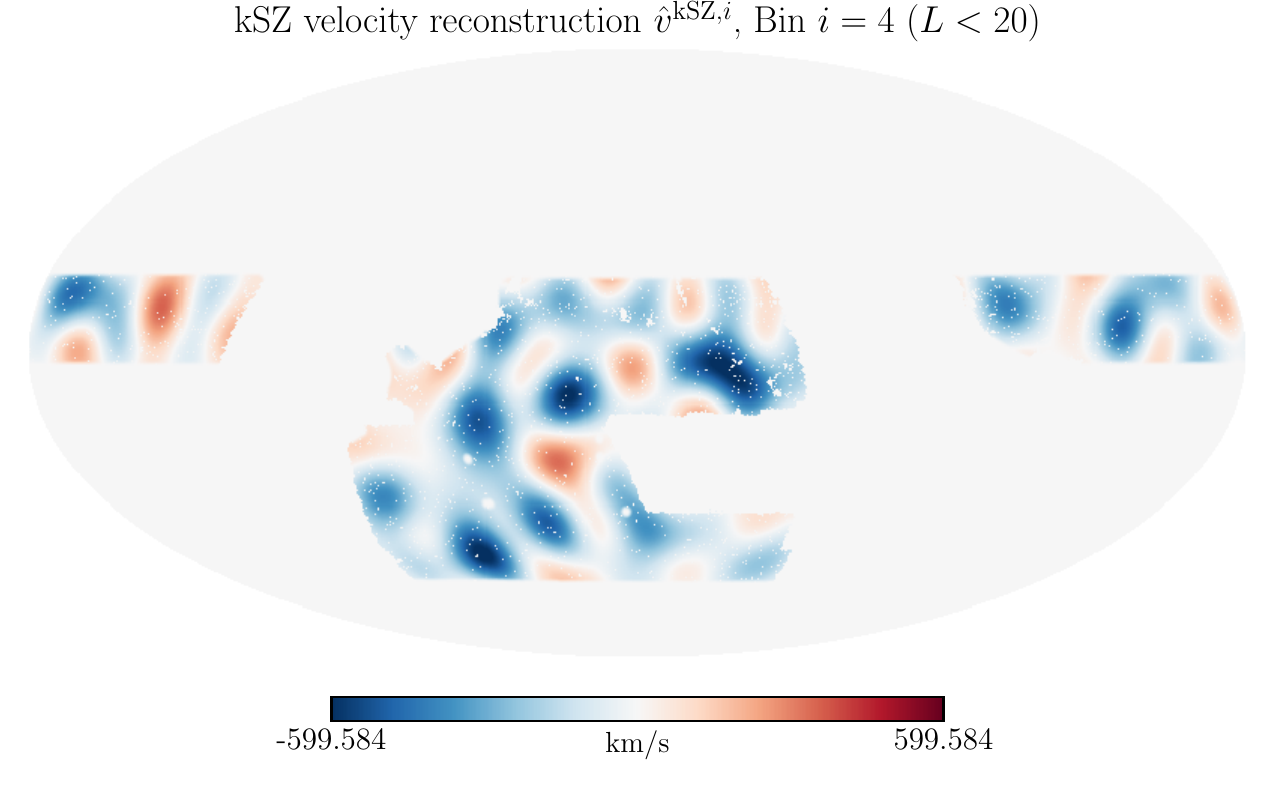}
\caption{The reconstructed velocity maps in each bin, filtered such that only the $L<20$ modes are shown. See Appendix~\ref{app:reconmaps} for maps constructed with other filters.}\label{fig:velocity_reconstructed_maps_l20}
\end{figure*}

\subsection{kSZ velocity reconstruction pipeline}~\label{sec:kszvelrecon}

\subsubsection{kSZ-galaxy cross-correlation}

The kSZ temperature anisotropy $\Delta T^{\rm {kSZ}}$ takes the form
\begin{align}
\frac{\Delta T^{\rm {kSZ}}(\hat n)}{\bar T} = &- \sigma_T\int d\chi  a(\chi)  v_r (\chi,\hat n) n_e(\chi,\hat n) ,\\
\equiv& - \int d\chi    v_r (\chi,\hat n)\dot \tau(\chi, \hat n) 
\end{align}
where $\bar T$ is the mean temperature; $\sigma_T$ is the Thomson scattering cross section; $a(\chi)$ is the scale factor at comoving distance $
\chi$; $v_r(\chi, \hat n) \equiv \vec v \cdot \hat n$ is the radial velocity at $(\chi,\hat n)$; and $n_e(\chi,\hat n)$ is the electron density at $(\chi,\hat n)$. We encompass several of these quantities into the definition of differential optical depth $\dot \tau (\chi, \hat n)$ above.

Due to the large-scale radial velocity modulation, the kSZ-galaxy cross correlation 
$\left<\Delta T^{\mathrm{kSZ}}_{\ell m} \delta^{g^i}_{\ell^\prime m^\prime}\right>$ is statistically anisotropic (ie, not proportional to $\delta _{\ell \ell^\prime} \delta_{m m^\prime}$, where $\delta_{XY}$ is the Kronecker delta). In particular, 

\begin{align}
 \left<\frac{\Delta T^{\mathrm{kSZ}}_{\ell m} }{\bar T}\delta^{g^i}_{\ell^\prime m^\prime}\right> =  \sum_{\ell_1 m_1 \ell_2 m_2}&W^{\ell_1 \ell_2 \ell}_{m_1 m_2 -m}\times\nonumber
 \\&\int d\chi \left<v_r{}_{\ell_1 m_1}(\chi)\dot \tau  _{\ell_2 m_2} (\chi)\delta^{g^i}_{\ell^\prime m^\prime}\right>,\label{generalbispectrum}
\end{align}
 where $W^{\ell_1 \ell_2 \ell_3}_{m_1 m_2 m_3}$ is a mode-coupling matrix of the form:
\begin{align}
W^{\ell_1 \ell_2 \ell_3}_{m_1 m_2 m_3} = \sqrt{\frac{(2 \ell_1+1)(2 \ell_2+1)(2 \ell_3+1)}{4\pi}}&\wignerJ{\ell_1}{\ell_2}{\ell_3}{0}{0}{0}\times\nonumber\\
&\wignerJ{\ell_1}{\ell_2}{\ell_3}{m_1}{m_2}{m_3}.
\end{align}
To probe large scales in the radial velocity, we can work in the squeezed limit, where where $\ell_1\equiv L\ll \ell_2, \ell_3$. In that limit, the three-point function in Equation~\eqref{generalbispectrum} can be simplified according to 
\begin{align}
\left< v_r{}_{LM}(\chi) \dot\tau_{\ell_2 m_2}(\chi)\delta^{g^i}_{\ell^\prime m^\prime}\right>&\approx v_r{}_{LM}(\chi)\left<\dot\tau_{\ell_2 m_2}(\chi)\delta^{g^i}_{\ell^\prime m^\prime}\right>\\
&\equiv v_r{}_{LM}(\chi) \left( C_{\ell^\prime}^{\dot\tau {g^i}}(\chi)\delta _{\ell^\prime \ell_2}\delta_{m^\prime m_2}\right),
\end{align}
where the form of the last line results from the statistical isotropy of $\left<\dot \tau_{\ell_2 m_2}(\chi)\delta^{g^i}_{\ell^\prime m^\prime}\right>$. 
Using this approximation simplifies Equation~\eqref{generalbispectrum} to
\begin{align}
\left<\frac{\Delta T^{\mathrm{kSZ}}_{\ell m} }{\bar T}\delta^{g^i}_{\ell^\prime m^\prime}\right> =     \int d\chi \sum_{LM} (-1)^{M}\wignerJ{\ell}{\ell^\prime}{L}{m}{m^\prime}{-M}     f_{\ell  \ell^\prime L}\times \nonumber\\
 C_{\ell^\prime}^{\dot\tau {g^i}} v_r{} _{LM}(\chi)
\end{align}
where 
\begin{equation}
f_{\ell  \ell^\prime L} =  \sqrt{\frac{(2\ell+1)(2\ell^\prime+1)(2L+1)}{4\pi}}\wignerJ{\ell}{\ell^\prime}{L}{0}{0}{0}.
\end{equation}
{This demonstrates that the mode coupling between $T$ and $\delta^g$ is proportional to the projection of $v_r$. we can use this property to write down a quadratic estimator with $T$, $\delta^g$ that uses this mode coupling to reconstruct $v_r$.}

\subsubsection{Harmonic-space quadratic estimator for $v_r$}

~\cite{2018PhRvD..98l3501D,2020PhRvD.102d3520M,2023JCAP...02..051C,2024arXiv240500809B} show that the minimum variance, unbiased quadratic estimators of a redshift-integrated velocity field  $v^i_r{}_{LM}$ are
\begin{equation}
\hat v_r^{i}{}_{LM} = A_L^{i}\sum_{\ell m; \ell^\prime m^\prime}(-1)^M \wignerJ{\ell}{\ell^\prime}{L}{m}{m^\prime}{-M}  f_{\ell \ell^\prime L}C_{\ell^\prime}^{\tau {g^i}}\tilde{T}_{\ell m}{\tilde{\delta ^{g^i}}_{\ell^\prime m^\prime}},\label{estimator_harmonic}
\end{equation}
where $\tilde X_{\ell m}\equiv\frac{X_\ell m}{C_\ell^{XX}}$ indicates the multipole moments of an inverse-variance-filtered field (where $C_\ell^{XX}$ includes signal and noise); $C_\ell^{\tau g^i}\equiv\int d\chi C_\ell^{\dot \tau g^i}(\chi)$; and

\begin{equation}
A_L^{i} = (2L+1)\left(\sum_{\ell;\ell^\prime}\frac{\left(f_{\ell \ell^\prime L}C_{\ell^\prime}^{\tau {g^i}}\right)^2}{C_\ell^{TT}C_{\ell^\prime}^{g^i g^i}}\right)^{-1}.\label{AL_def}
\end{equation}
Note that $T_{\ell m}$ is an estimate of $T^{\mathrm{kSZ}}_{\ell m}$, and includes the signal as well as other sources of foregrounds and noise (such as the primary CMB).
The projected quantity that we estimate $v^i_r{}$  {is related to the three-dimensional velocity $v_r(\chi)$} by the following integral:
\begin{equation}
v^i_r{}_{LM} \approx \int d\chi \frac{C_{\ell=\bar \ell}^{\dot \tau g^i}(\chi)}{C_{\ell=\bar \ell}^{\tau g^i}}v_r{}_{LM}(\chi),
\end{equation}
where $\bar \ell$ is a reference multipole that contributes most of the signal-to-noise to the measurement.
Due to the slow redshift evolution (compared to the galaxy window functions) of the electron distribution over the redshift bins in which the galaxies are binned, using 
\begin{align}
v^i_r{}_{LM}\approx\int d\chi W^i(\chi)v_r{}_{LM}(\chi),
\end{align}
where $W^i(\chi)$ is the redshift distribution ($\frac{dN}{d\chi}$) of the galaxy window used in the estimation, computes the velocity models accurately to within a few percent.

\begin{figure}
\includegraphics[width=\columnwidth]{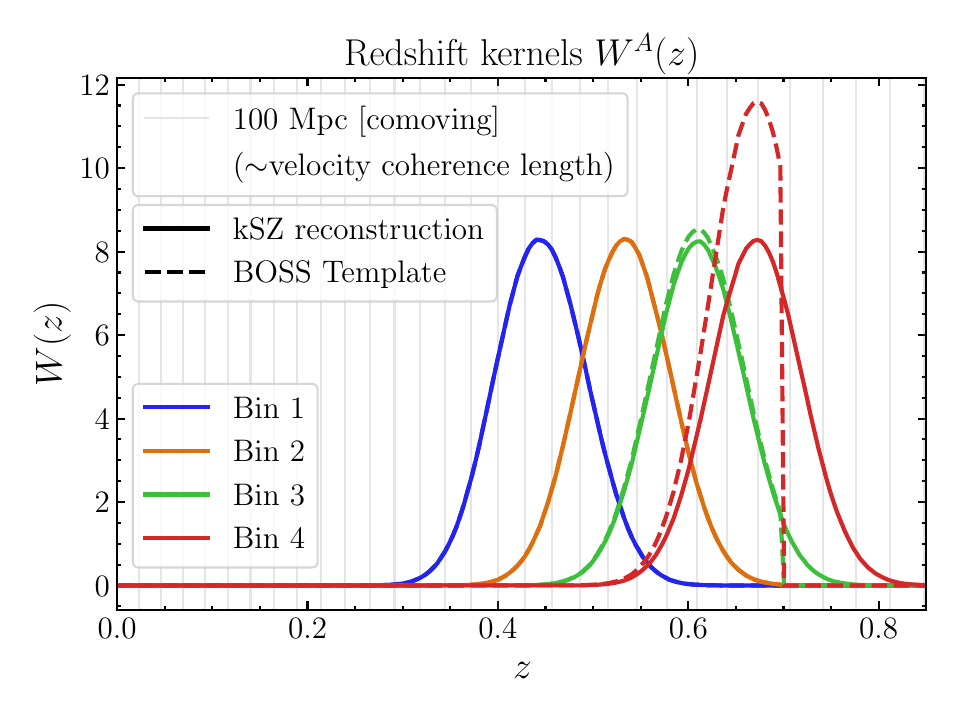}
\caption{Redshift kernels used in the cross-correlation modeling for the kSZ reconstruction on the DESI LRGs (solid lines), and for the BOSS templates (dotted lines). The BOSS and DESI bins are equal (up to normalization) except for the  $z<0.7$ cut, imposed to reflect the redshift coverage of the survey. All distributions are normalized to integrate to 1. Note the lower redshift BOSS and DESI bins coincide.}\label{fig:redshift_dists}
\end{figure}

\subsubsection{Real-space quadratic estimator for $v_r$}

In practice, we use a real-space version of Equation~\eqref{estimator_harmonic}. This estimator is advantageous as it requires only forward-and-backward spherical harmonic transformations on the data (and no further complicated computations such as of Wigner-$J$ symbols, except in the computation of the normalization $A_L$) . The real-space estimator is
\begin{equation}
\hat v^{i}_r{}_{LM}= A^{i}_L \left( \tilde T(\hat n) \zeta^{i}(\hat n)\right)_{LM},
\end{equation}
where $\zeta^{i}(\hat n)$ is a filtered version of a galaxy field. The filtering is such that it has the (modeled) scale dependence of the $\tau$ field:
\begin{equation}
\zeta^{i}_{l m} =  {C_\ell^{\tau g^i}}\tilde{\delta ^{g^i}}_{\ell m}
\end{equation} 
where $\tilde \delta^g$ is an inverse-variance-filtered galaxy overdensity field. The real-space estimator $\hat v^{i}_r{}_{LM}$ is the spherical-harmonic transform of their real-space product, filtered by $A^i_L$.

\begin{figure}
\includegraphics[width=\columnwidth]{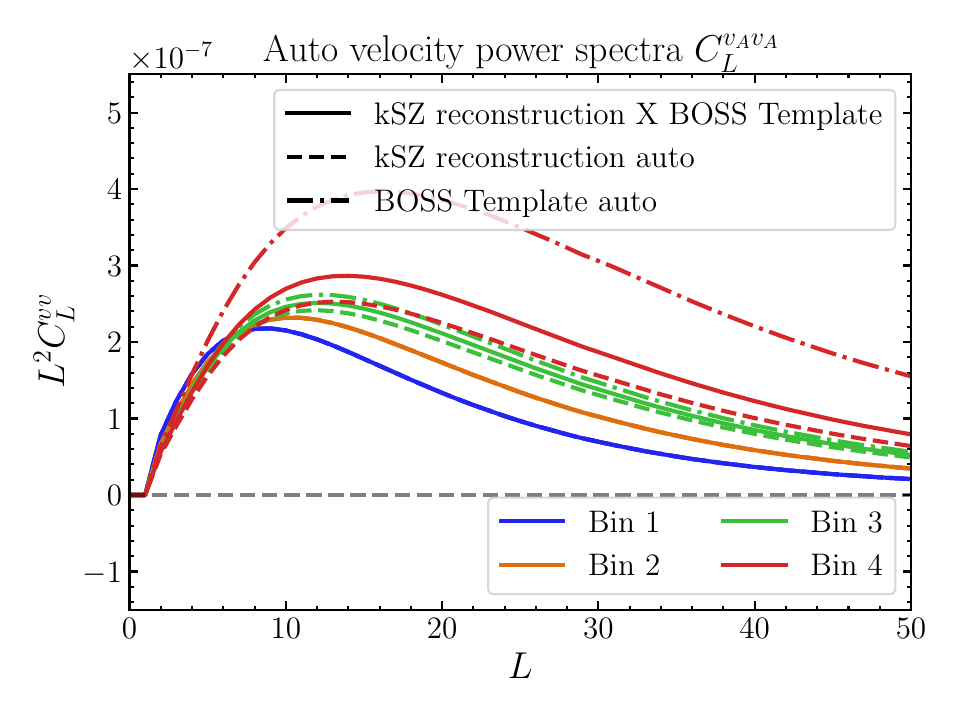}
\caption{The theory auto-power-spectra for the four redshift bins, as computed with \texttt{ReCCO}. We show the expected signal for the auto-spectra of the reconstructions and of the templates, as well as their cross power spectra. {Note that for Bins 1 and 2, the BOSS redshift bins overlap with the kSZ redshift bins such that the solid, dashed, and dotted lines all lie on top of each other for these bins. The sharp cutoff at $z>0.7$ in the BOSS template leads to the separation of the curves seen for Bin 4 and, to a lesser extent, Bin 3.}}\label{fig:autopower_theory}
\end{figure}

\begin{figure*}
\includegraphics[width=0.49\textwidth]{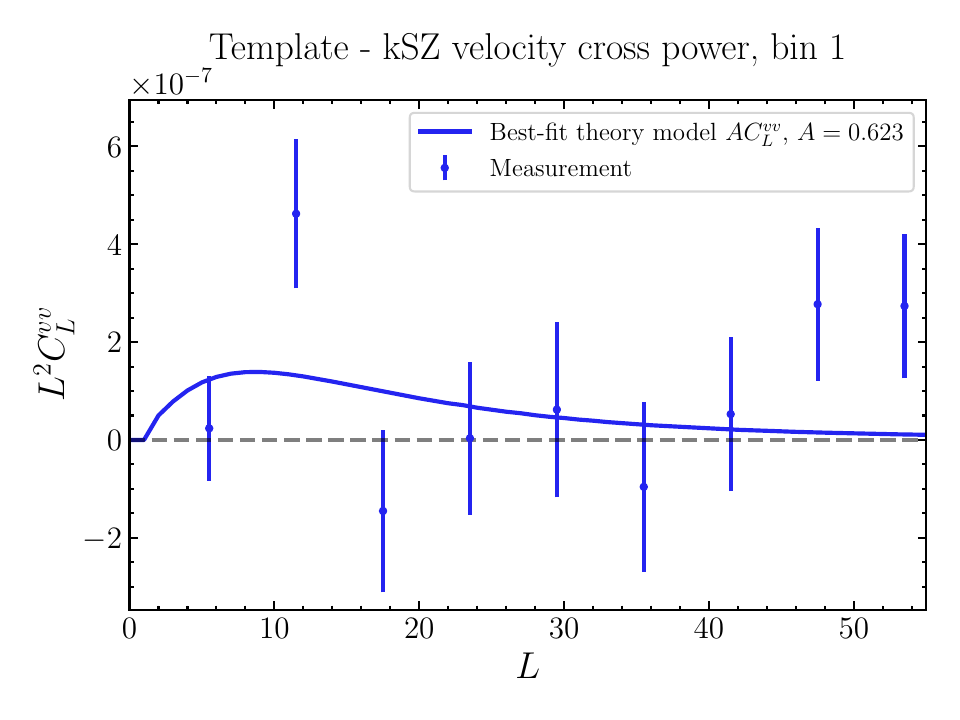}
\includegraphics[width=0.49\textwidth]{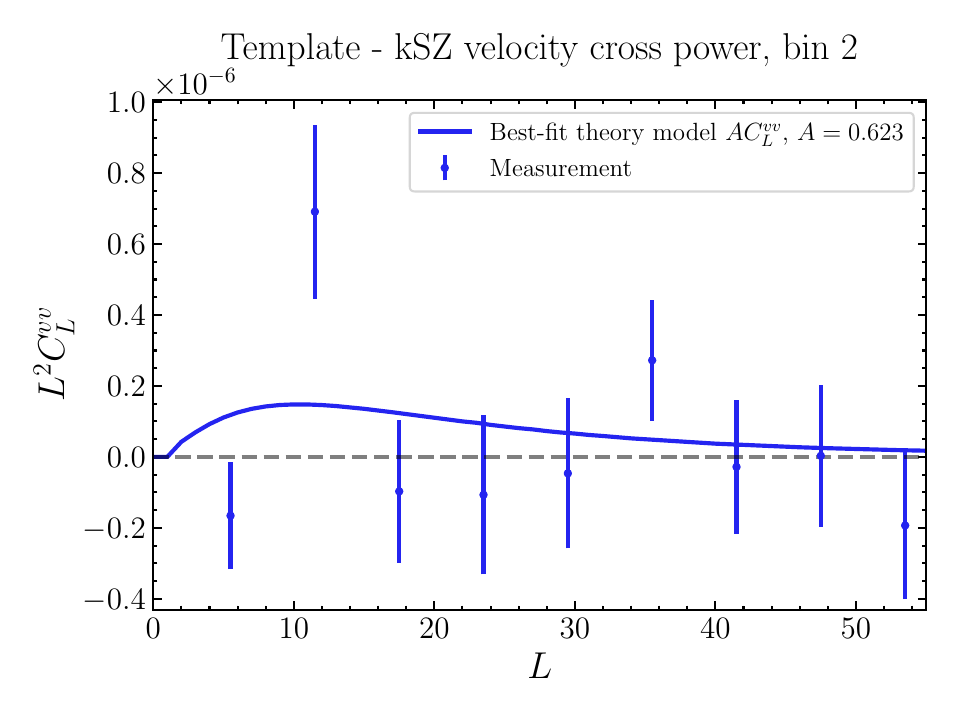}
\includegraphics[width=0.49\textwidth]{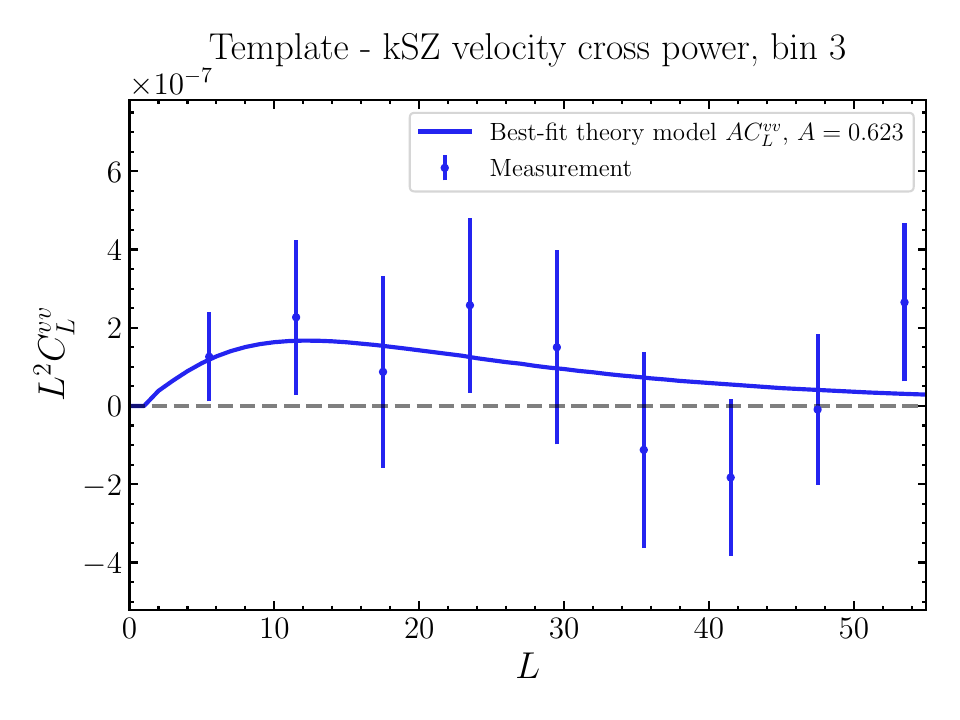}
\includegraphics[width=0.49\textwidth]{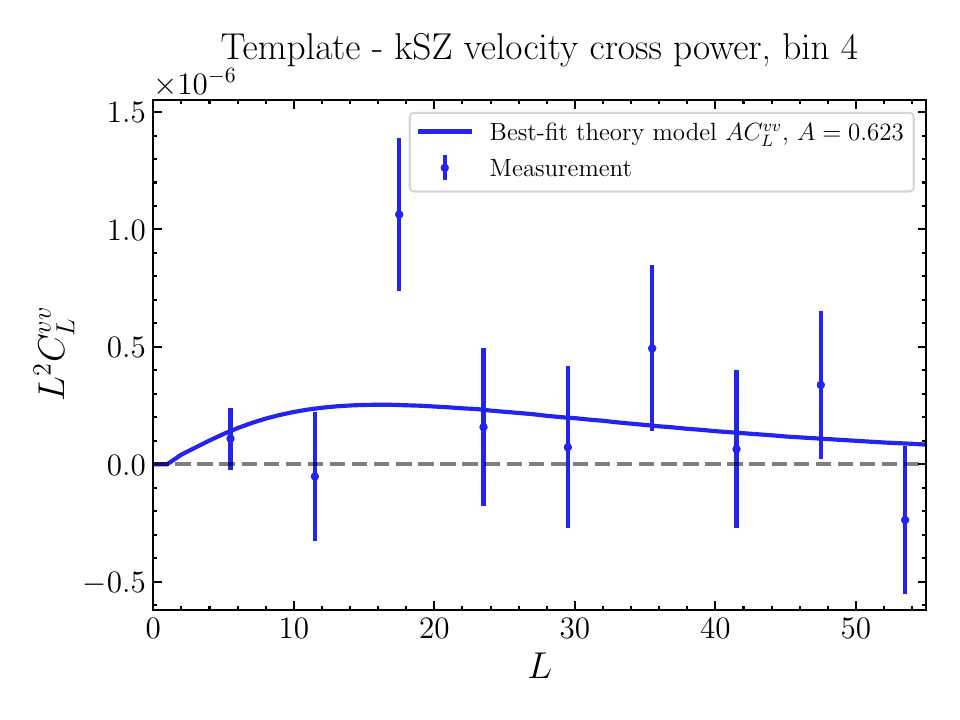}
\caption{The measured cross power spectra of the four velocity templates and reconstructions, along with the best-fit models.}\label{fig:bestfit_and_Data}
\end{figure*}
\subsubsection{Inputs to the estimator and the optical depth bias}

The estimators depend on the multipole moments of the temperature field $T_{\ell m}$; the multipole moments of the galaxy overdensity field $\delta^{g^i}_{\ell m}$;  their auto power spectra $C_\ell^{TT}$, $C_\ell^{g^ig^i}$; and the galaxy-electron cross power spectrum $C_\ell^{\tau g^i}$. The auto power spectra $C_\ell^{TT}$ and $C_\ell^{g^ig^i}$, which are used for inverse-variance-filtering, are the auto spectra of the signal including all noise contributions and are estimated directly from the data.

In contrast, $C_\ell^{\tau g^i}$, which quantifies how the galaxies are distributed with respect to the electrons, is not obtained from the data and requires a model. We describe the model we use in Appendix~\ref{app:cltaug}. An incorrect model will lead to a biased (mis-normalized) estimator---this is the  ``{optical depth bias}'' such that 
\begin{equation}
\hat v^{i}_r{}_{LM} = b_L ^{i} v^{i}_r{}_{LM}.
\end{equation} 
where $b_L ^{i}$ is the bias in question.
 $b^i_L$ can be expressed in terms of the model error on $C_\ell^{g\tau}$ and is scale-independent on the scales $L$ of interest, so there are in principle only $i$ optical-depth-bias parameters. In practice, we will reduce this to one parameter (which we will label $A$), as we do not have the signal-to-noise to detect any redshift evolution in our model.

The filters used are presented in more detail in Appendix~\ref{app:filters}.

\subsubsection{Results of the reconstruction: radial velocity maps}

We show  the reconstructed velocity fields in Figure~\ref{fig:velocity_reconstructed_maps_l20}. As  small-scale noise fluctuations dominate, we show  the maps filtered to preserve only the $L<20$ information.  We show in Appendix~\ref{app:reconmaps} the same plots but with only the $L<10$ and $L<50$ information shown.

\subsection{Cross-correlation with an external velocity template}

\subsubsection{Velocity power spectrum measurement}\label{sec:clvv_estimate}

After creating estimates of $\hat v_r^i$ we then estimate the velocity power spectrum $C_\ell^{vv}$ by cross-correlating the kSZ-reconstructed $\hat v_r^{\mathrm{kSZ},i}$ with the velocity templates created by performing continuity-equation velocity reconstruction on the BOSS galaxies $\hat v_r^{\mathrm{template},I}$. {We refer to our estimate as $C_\ell^{\hat v_r^{{\rm{kSZ}},i}\hat v_r^{{\rm{template}},J}}$.}  We perform the power spectrum measurements with \texttt{pymaster}~\citep{2019MNRAS.484.4127A}.\footnote{\url{https://namaster.readthedocs.io}} This is a python implementation of the MASTER algorithm~\citep{2002ApJ...567....2H}, which estimates a multipole-binned mask-decoupled power spectrum of two masked Gaussian fields. We note that this is suboptimal to, say, a quadratic maximum likelihood
approach~\citep{2001PhRvD..64f3001T}, which we leave for future work. 

While there are 16 independent cross-correlations (as $i=1,2,3,4$ and $J=1,2,3,4$), we only measure the four ``intra-bin'' power spectra where $i=J$. This avoids bias from modelling uncertainties, in particular in the tails of the redshift distributions.

\subsubsection{Velocity covariance matrix}\label{sec:covmat_estimation}
We calculate the covariance matrix $\mathbb {C} \equiv\mathrm{Cov}(\hat C_L^{\hat v_r^{{\rm{kSZ}},i}\hat v_r^{{\rm template},I}},\hat C_{L^\prime}^{\hat v_r^{{\rm{kSZ}},j}\hat v_r^{{\rm template},J}})$ by creating 3200 null Gaussian simulations of the NILC temperature map and the DESI galaxy overdensity maps. We do this by drawing random (full-sky) realizations of {a temperature field} with the same power spectrum as the NILC temperature map, and {galaxy fields with} the same power spectra as the tomographically binned DESI galaxy samples $C_\ell^{g^ig^j}$ (including $i\ne j$); this is performed easily with \texttt{healpy}'s \texttt{synalm()} function. {We include no correlation between the simulated galaxies and temperature, and so expect to recover a null signal.} We  mask these with the appropriate masks and then run our full velocity reconstruction pipeline on these simulations (as described in Section~\ref{sec:kszvelrecon}).

\begin{figure*}
\includegraphics[width=0.49\textwidth]
{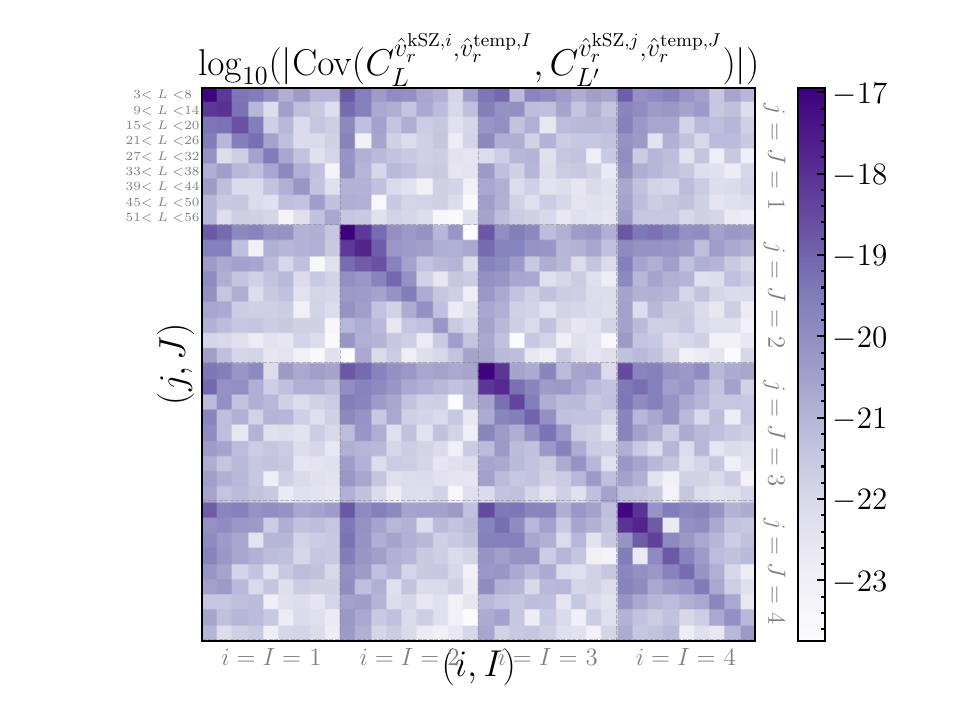}
\includegraphics[width=0.49\textwidth]{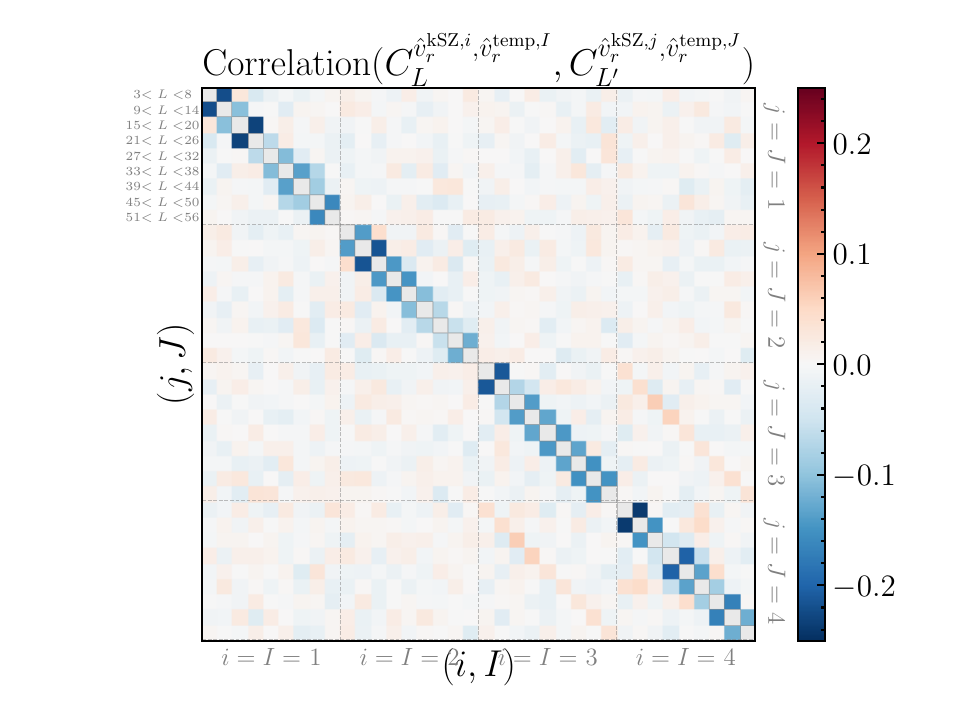}
\caption{\textit{Left}: the log of the absolute value of the covariance matrix for the intra-bin analysis. \textit{Right}: the correlation matrix for the intra-bin analysis. The correlation matrix is defined here as the covariance matrix divided by the square root of its diagonal on both axes. We have removed the diagonal from the visualization, as it is by definition equal to 1.}\label{fig:covmat_intrabin}
\end{figure*}

{In practice, we find that these do not adequately explain the entire variance of the velocity measurements. While the simulations match the data at the 2-point level, we find that there is a mismatch of order $1.2^2$ at the 4-point level. In particular, the measured auto power spectra of the velocities recpmstricted from the simulations agree with the theoretical calculation}
\begin{equation}
N_L^{(0)}{}^{\hat v_r^{ \mathrm{kSZ},i}\hat v_r^{\mathrm{kSZ},i}} = A_L^i,
\end{equation}
{where $A_L^i$ is the same as the amplitude in Equation~\eqref{AL_def}; this is expected. However, due to an unmodelled 4-point function in the data, we find that the data has an excess of order $1.2^2$ (ie, $N_L^{(0)}{}^{\hat v_r^{ \mathrm{kSZ},i}\hat v_r^{\mathrm{kSZ},i}} \approxeq 1.2^2A_L^i,$). This may be due to higher order $\left<ggTT\right>$ biases not included in the simulations; a similar phenomenon was found in~\cite{2024arXiv240500809B}. To correct for this, we simply multiply the simulations by $1.2$, which results in the auto power spectra of the simulated velocity reconstructions matching  those of the data. This only has the effect of increasing the errorbars by a factor of 1.2, and does not affect the data or modelling.}

We then measure $\hat C_L^{\hat v_r^{{\rm{kSZ_{\mathrm{sim}}}},i}\hat v_r^{{\rm{template}},I}}$ (as described in Section~\ref{sec:clvv_estimate}), using the \textit{true} velocity field from the BOSS galaxies. We verify that the mean measurement is zero, and use the covariance of these measurements as our covariance matrix. 

\subsection{Likelihood for detection}

We write a likelihood
\begin{align}
-2\ln \mathcal {L} (\hat C_L^{\hat v_r^{{\rm{kSZ,}}i}\hat v_r^{{\rm{template}},I}},A) &= \left(\hat C_L^{\hat v_r^{{\rm{kSZ,}}i}\hat v_r^{{\rm{template}},I}}- AC_L^{v^{i}v^I} \right ) \mathbb{C}^{-1} \nonumber\\
&\hspace{2em}\cdot\left(\hat C_{L^\prime}^{\hat v_r^{{\rm{kSZ}},j}\hat v_r^{{\rm{template}},J}}- AC_{L^\prime}^{v^{j}v^J} \right )\label{likelihood_def}\\
&\equiv \chi^2\label{chi2_def}
\end{align}
where $C_L^{v^iv^J}$ (with no hat) indicates a theory model. $A$ is an ampltidue parameter that encompasses the uncertainty in the electron and galaxy model: the optical depth bias. As we have low signal to noise, we jointly fit only one parameter across all redshift bins, although we will explore redshift dependence of $A$ in Section~\ref{sec:redshift_dependence}. We maximize the likelihood with respect to $A$ to find a best-fit amplitude for the theory model. We then use 6800 more Gaussian simulations, which are created the same way as those we used to estimate the covariance matrix (ie, as described in Section~\ref{sec:covmat_estimation}), but which, importantly, are {independent from those used in the covariance estimation}, and maximize the likelihood with respect to $A$ for all of these simulations. We verify that the mean is zero, and use the distribution of measurements to quantify the uncertainty (covariance) on $A$.

\section{Theory modelling}

We require a theory model for two quantities: the electron-galaxy angular cross-correlation power spectrum, $C_\ell^{\tau g^i}$, and the velocity auto-power spectrum, $C_\ell^{v^A v^B}$. We use \texttt{class\_sz}\footnote{\url{https://github.com/CLASS-SZ/class_sz}}~\citep{2023arXiv231018482B,2023JCAP...03..039B} to model $C_\ell^{\tau g^i}$, and our modified version of \texttt{ReCCO} to model $C_L^{v^A v^B}$. We present the model for the velocity power spectrum in this Section, and we present the model for  $C_\ell^{\tau g^i}$ in Appendix~\ref{app:cltaug}.

\begin{figure*}
    \centering
    \includegraphics[width=0.32\textwidth]
    {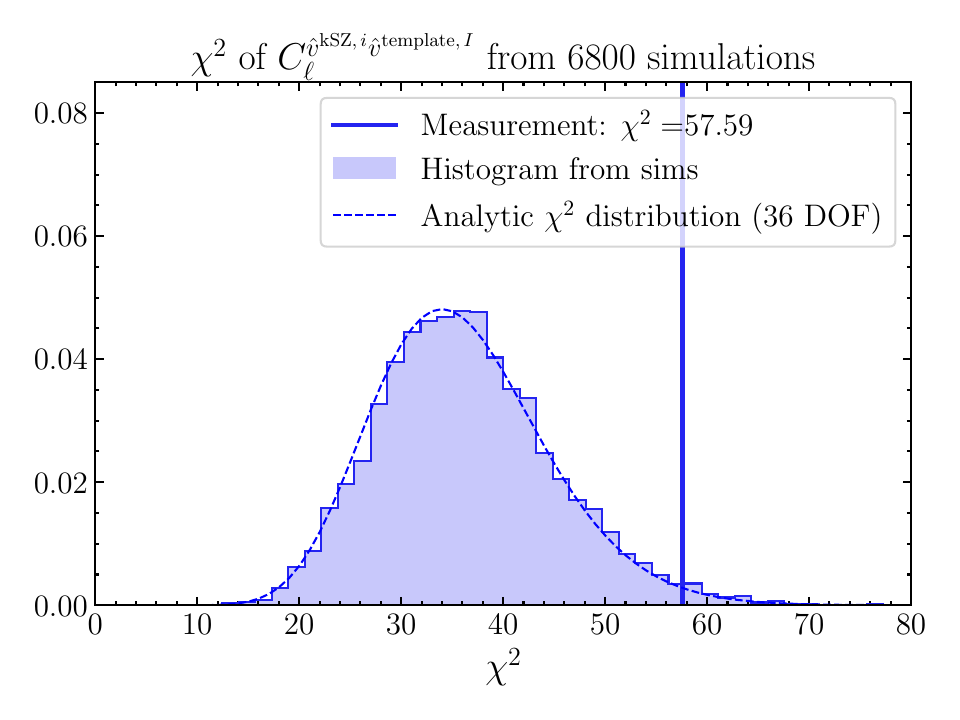}
    \includegraphics[width=0.32\textwidth]{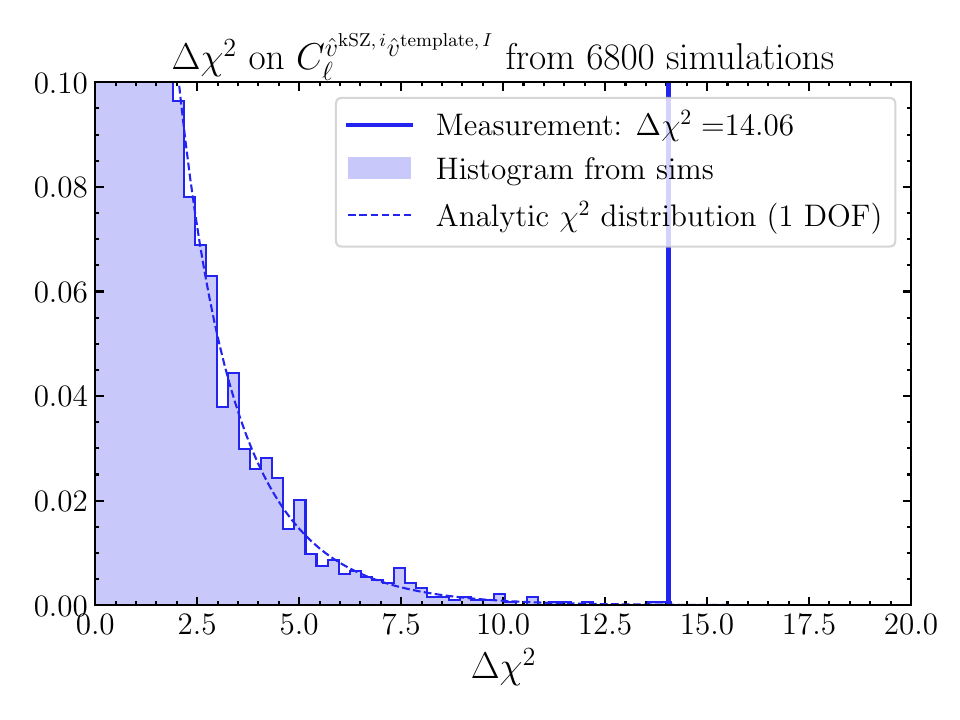}
    \includegraphics[width=0.32\textwidth]{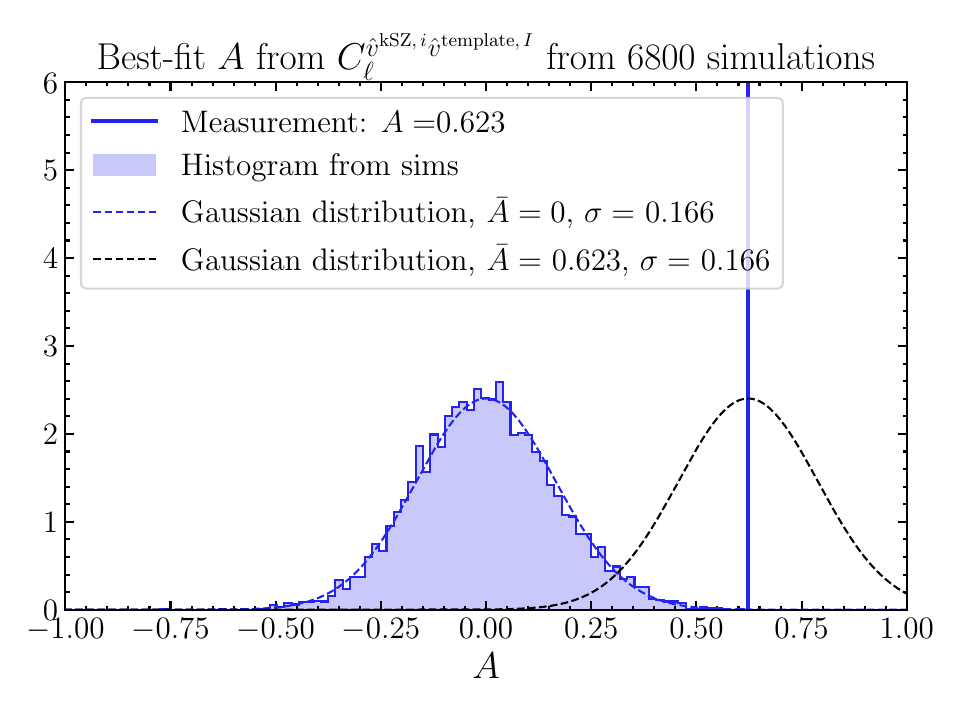}

    \caption{\textit{Left:} The distribution of the $\chi^2$ with respect to zero of $C_L^{v^iv^I}$from 6800 simulations, with the $\chi^2$ of the measurement also indicated. \textit{Center}: The distribution of the $\Delta\chi^2$, ie the difference between the $\chi^2$ with respect to the best fit and the $\chi^2$ with respect to zero of $C_L^{v^iv^I}$from 6800 simulations, with the $\Delta\chi^2$ of the measurement also indicated. Only two of the 6800 simulations have a higher $\Delta\chi^2$ than the measurement. 
    \textit{Right}: The distribution of the recovered best-fit values of $A$,
    for the simulations and with the  data measurement indicated. In all cases the histograms are normalized to be density distributions, and similarly the Gaussian plots have unit normalization.
    }
    \label{fig:chi2sdistribution_auto_fitfacs}
\end{figure*}

\subsection{Radial velocity power spectrum from \texttt{ReCCO}}

We model the radial velocity power spectrum $C_L^{v^Av^B}$ on large scales  within the linear $\Lambda$CDM model by modifying code previously implemented in \texttt{ReCCO}.  $C_L^{v^Av^B}$ is modelled as
\begin{align}
C_L^{v^A v^B} =& \int d\chi_1 d\chi_2 W^A(\chi_1) W^B(\chi_2)\nonumber\\
&\times\int \frac{k^2 dk}{(2\pi)^3}\mathcal K_L^v(\chi_1, k)\mathcal  K_L^v(\chi_2, k)P_{\mathrm{lin}}(\chi_1,\chi_2,k),\label{general_clvv}
\end{align}
where $W^X(\chi)$ are the redshift kernels of the velocity fields we are interested in, {and the integral over the 3d Fourier-space vector $k$ calculates how the gravitational impact of the large scale matter distribution induces velocities. Here} $P_{\mathrm{lin}}(\chi_1,\chi_2,k)$ is the  linear matter power spectrum, and {the velocity kernel} $\mathcal K_L^v(\chi, k)$ is constructed from the velocity transfer functions.  

The relevant redshift kernels for $C_\ell^{v^iv^I}$ (which appears in our likelihood) are $\frac{dN^i}{d\chi}$  and $\frac{dN^I}{d\chi}$, the redshift distributions for the four DESI bins and BOSS templates respectively. These are shown in Figure~\ref{fig:dndz_4bins}.

For $W^i(\chi)$ (relevant for the kSZ-reconstructed velocity), we use:
\begin{align}
W^{i,\mathrm{kSZ}}(\chi) = \frac{\frac{dN^i}{d\chi}}{\int\frac{dN^i}{d\chi} d\chi }
\end{align}
where $\frac{dN^i}{d\chi}$ is the redshift distribution of the tomographic DESI galaxy bin we used in the kSZ velocity reconstruction for bin $i$. For the BOSS templates, which have been reweighted to match the same redshift distribution but which contain no information from galaxies at $z>0.7$, we use 
\begin{align}
W^{I,\mathrm{template}}(\chi) = \begin{cases}
\dfrac{\frac{dN^I}{d\chi}}{\int_{z<0.7}\frac{dN^I}{d\chi}d\chi} & z<0.7\\
0& z\geq0.7,
\end{cases}
\end{align}
where $\frac{dN^{I}}{d\chi}$ are the same as $\frac{dN^i}{d\chi}$.
These are shown in Figure~\ref{fig:redshift_dists}. Note the different normalization in the highest redshift bins, which is due to the fact that all redshift kernels are normalized to integrate to 1.

{The velocity kernel has the form}
\begin{equation}
\mathcal K_L^v(\chi, k) = 4 \pi i^L\frac{f(\chi) H(\chi) a(\chi)}{(2L+1)k}\left(L j_{L-1}(k\chi) - (L+1) j_{L+1}(k\chi)\right),
\end{equation}
{where $j_L(x)$ is the Bessel function of degree $L$.}
Note that this only includes the contribution to the kSZ signal from the ``local Doppler'' term (ie, the relative velocity of a remote object with respect to us---the part sourced by the gravitational field around it), and neglects the signal induced by the intrinsic CMB dipole that an object sees.\footnote{On very large scales, these contributions may be relevant to the auto power spectrum of the kSZ-reconstructed velocity, and contain useful information to constrain primordial physics (see, e.g.,~\citealt{2020PhRvD.101l3508C}), but they do not appear in the velocity field template that was reconstructed from the continuity equation and so are not relevant for our signal.}

At large $L$, and when $W(\chi)$ has wide support (ie, large redshift bins), the Limber approximation~\citep{1953ApJ...117..134L} allows Equation~
\eqref{general_clvv} to be simplified to the very familiar
\begin{equation}
C_L^{v^Av^B}=\int \frac{d\chi}{\chi^2} W^A(\chi) W^B(\chi) \left(\mathcal K_L^v\left(\chi,\frac{L+1/2}{\chi}\right)\right)^2 P_{\mathrm{lin}}\left(\chi,\frac{L+1/2}{\chi}\right).
\end{equation}
On large scales, this is not a good approximation. Thus, \texttt{ReCCO} uses the beyond-Limber corrections of~\cite{2020JCAP...05..010F} to compute the power spectra. The implementation is described in detail in~\cite{2023JCAP...02..051C}.

The theoretical signal is shown in Figure~\ref{fig:autopower_theory}. We show the power spectra calculated both with the window functions appropriate for the kSZ reconstruction (``kSZ reconstruction auto'', $C_L^{v^iv^i}$) and for the BOSS templates (``BOSS template auto'', $C_L^{v^Iv^I}$) as well as one of each (``kSZ reconstruction X BOSS template'', $C_L^{v^iv^I}$), which we will measure by taking the cross-spectra of the reconstruction with the templates. The models differ for the higher-redshift bins due to the sharp cut-off in $z$ of the BOSS templates at $z=0.7$.

\begin{figure*}
    \centering
    \includegraphics[width=1.3\columnwidth]{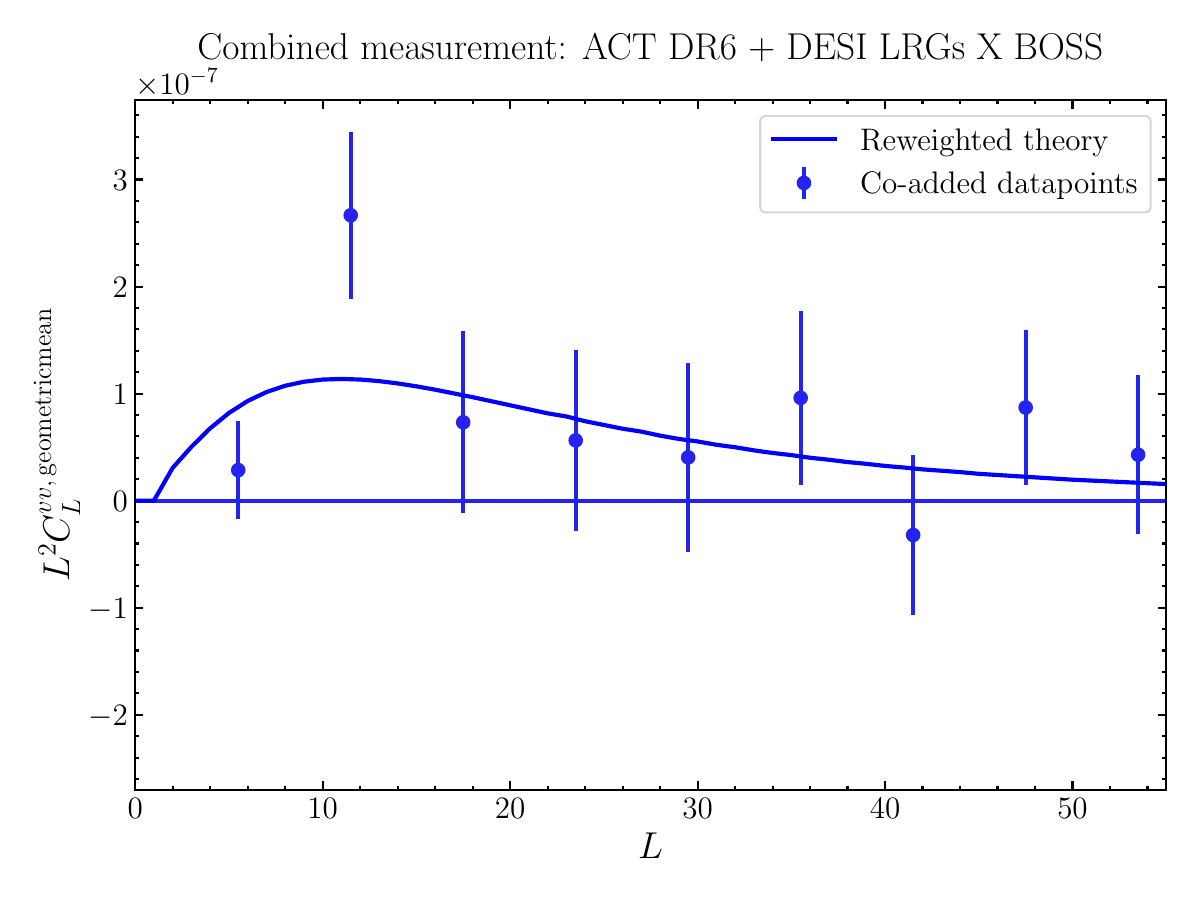}
    \caption{The reweighted, co-added datapoints, and the reweighted theory. This plot is shown for visualization purposes; in practice, the datapoints of Figure~\ref{fig:bestfit_and_Data} are analyzed.}
    \label{fig:reweighted_theory}
\end{figure*}

\section{Cross-correlation measurement and significance}

In this Section we present our measurements of the velocity power spectrum $C_L^{v^Av^B}$. All measurements are ``cross'' spectra, in that we are measuring a cross power spectrum between a continuity-equation-reconstructed template and a kSZ-reconstructed velocity. To avoid confusion, we refer to this spectrum (ie,  $C_L^{v^iv^I}$) as the ``intra-bin'' power spectra  (ie, as opposed to the ``auto-spectrum'', or the ``cross-spectrum'', which could more generically mean an inter-bin measurement of $C_L^{v^iv^J}$ for $i\ne J$\footnote{While we can measure $C_L^{v^iv^J}$, we do not model it as it is more sensitive to tails in the $\frac{dN}{dz}$ than the inter-bin measurement, and is thus difficult to model accurately without spectroscopically confirmed redshift distributions. However, the signal is expected to be non-zero, both due to long-range inter-bin correlations and the non-negligible redshift overlap of neighbouring bins. We leave the interpretations of such measurements to future work.}).

We measure the cross power-spectra with \texttt{pymaster}. In all cases, we use an $L$-binning scheme defined by linear bins with $\Delta L=6$ starting at $L_{\mathrm{min}}=3$ and up to $L_{\mathrm{max}}=56$ (note that we also include the bin $0<L<3$ in our \texttt{pymaster} binning scheme, measurement, and decoupling matrix, although we do not use it in the analysis). {We explore the sensitivity of the measurement to  $L_{\mathrm{min}}$ in Section~\ref{sec:scaledep}.}

\subsection{Datapoints and covariance matrix}

The measured ``intra-bin'' cross power spectra of the kSZ velocity reconstructions and the BOSS templates, $\hat C_L^{\hat v^{\mathrm{kSZ},i} \hat v^{\mathrm{template},I}}$ are shown in Figure~\ref{fig:bestfit_and_Data}. To measure their covariance matrix, we use 3200 of the simulations described above. This covariance matrix is shown in Figure~\ref{fig:covmat_intrabin}, along with the same object converted to a correlation matrix (ie, divided by the square root of its diagonal on both axes). It is clear here that the most important off-diagonal correlations are those between neighbouring multipole bins, although some structure in the correlations begin to be visible for the $L=L^\prime$ entries of $\mathrm{Cov}(C_L^{v_r^3v_r^3},C_{L^\prime}^{v_r^4v_r^4})$. This is  expected, as these redshift bins have a largest overlap, and the underlying signal $C_L^{v_r^{\mathrm{template},I}v_r^{\mathrm{template},J}}$ as well as the correlated reconstruction noise $N_L^{v_r^{\mathrm{kSZ},i}v_r^{\mathrm{kSZ},j}}$ are maximized here (recall there is no signal $C_L^{v_r^{\mathrm{kSZ},i}v_r^{\mathrm{kSZ},j}}$ or $C_L^{v_r^{\mathrm{kSZ},i}v_r^{\mathrm{template},J}}$ in the simulations, including for $i=J$).

\begin{figure}
\includegraphics[width=\columnwidth]{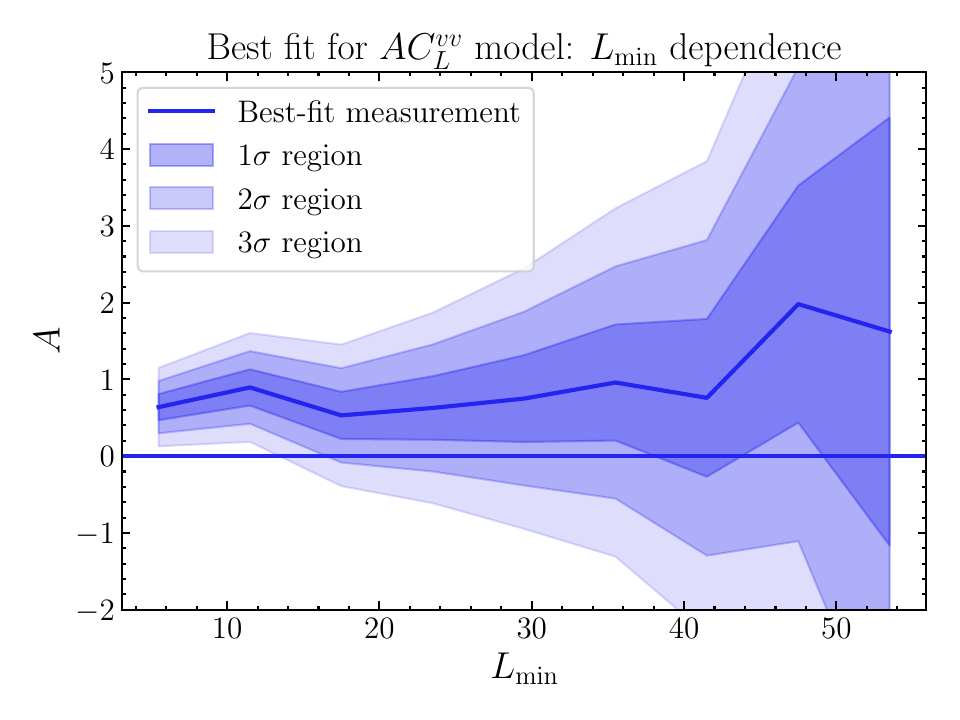}
\caption{The best-fit value of $A$ as a function of the minimum multipole included in the analysis $L_{\mathrm{min}}$. The 1$\sigma$, 2$\sigma$, and 3$\sigma$ regions are indicated. Much of the signal-to-noise comes from the largest scales, in particular the $3\sigma$ signal requires $L_{\mathrm{min}}\le12$, and a $2\sigma$ signal requires $L_{\mathrm{min}}\le20$.}\label{fig:minmultipole}
\end{figure}
\subsection{Consistency with zero}

The $\chi^2$ with respect to zero of our measurement is 57.59, corresponding to a PTE of 1.26\% (as calculated from an analytical $\chi^2$ distribution for 36 degrees of freedom, corresponding to four redshift bins with nine $L$ bins each). We show this, along with the distribution of $\chi^2$-to-zero from our null simulations, in Figure~\ref{fig:chi2sdistribution_auto_fitfacs}. Indeed, 114 (ie,  1.68\%) of the simulations have a higher $\chi^2$ than the measurement; this, along with the fact that the analytical $\chi^2$ is a good fit by-eye to the histogram on, indicates that the analytical $\chi^2$ distribution is a good description for these datapoints.

\begin{figure*}
\includegraphics[width=0.49\textwidth]{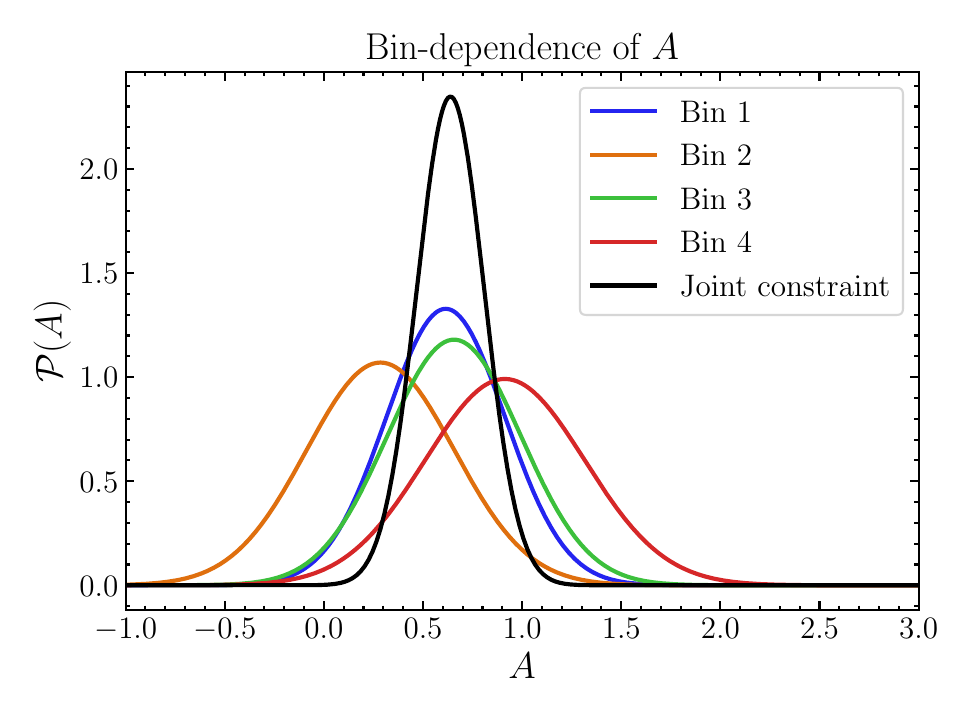}
\includegraphics[width=0.49\textwidth]{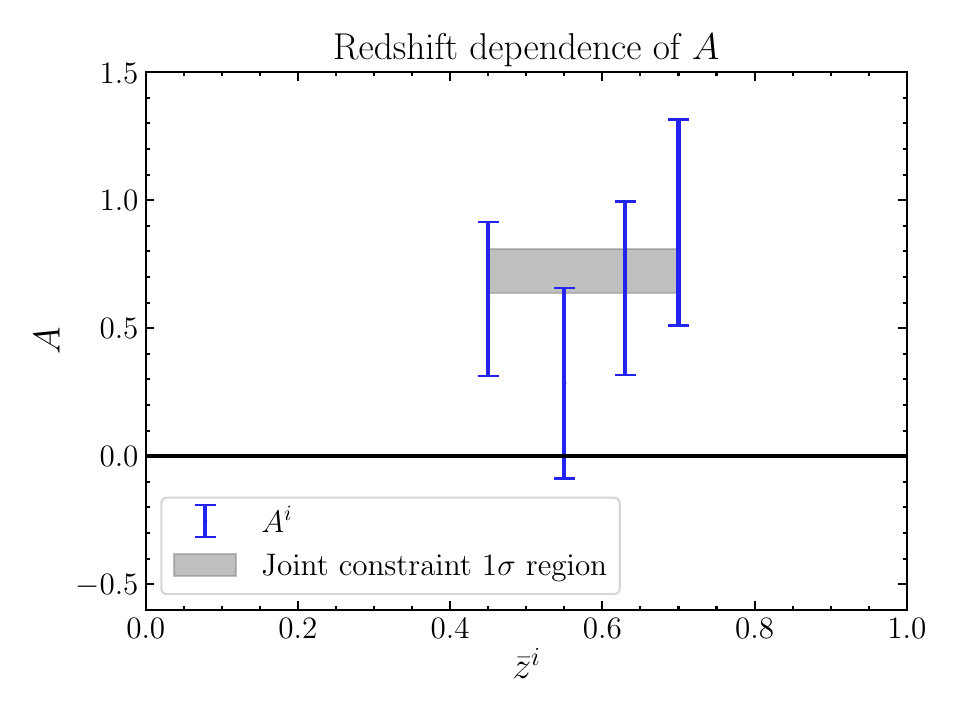}
\caption{The redshift dependence of $A$. \textit{Left:} Gaussian estimates of the posterior distributions of $A$ (assuming a uniform linear prior on $A$), which we define as $\mathcal P(A)\propto e^{-\frac{(A - A_{\mathrm{best fit}}^i)^2}{{2\sigma_i ^2}}}$, with $A_{\mathrm{best fit}}^i$ the measurement from the data and $\sigma_i$ the standard deviation of the best-fit of the simulations in that redshift bin (which are centred on zero). We also show the equivalent distribution for the joint best-fit, in black. \textit{Right:} The redshift dependence of $A^i$, plotted against the mean redshift of the bins (which we calculate by taking the $z$ at which $W^i(z)$ integrates to 1/2 from 0). The points are centred on the best-fit value of $A^i$, with the errorbar given by $\sigma_i$ as before. The $1\sigma$ region of the joint constraint is also indicated in gray.  }\label{fig:redshift_dependence}
\end{figure*}

\subsection{Measurement significance and recovered signal amplitude}

We maximize the likelihood defined in Equation~\eqref{likelihood_def} with respect to the amplitude parameter $A$, both for the data and also for our 6800 null simulations. The improvement in the $\chi^2$, ie $\Delta\chi^2\equiv \chi^2|_{0}-\chi^2|_{\mathrm{maximum-likelihood}}$, with $\chi^2|_{0}$ the $\chi^2$ when $A=0$ and $\chi^2|_{\mathrm{maximum-likelihood}}$ the $\chi^2$ at the maximum likelihood value,   is shown in Figure~\ref{fig:chi2sdistribution_auto_fitfacs}, in the centre panel. We overplot an analytic $\chi^2$ distribution for one degree-of-freedom. The improvement in the $\chi^2$ for the measurement is 14.06, corresponding to a PTE of 0.018\%---when comparing to the simulations, we note that this  $\Delta\chi^2$ is indeed greater than all but two (ie, 0.03\%)  of them. Given the agreement between the distributions and the analytical $\chi^2$ distributions, we can use either of these as a detection significance, and note that this corresponds to $\sim3.7\sigma$ evidence (from the simulations) or 3.9$\sigma$ from the analytical $\chi^2$ distribution. The $\chi^2$ with respect to the best fit is 43.53, corresponding to a PTE of 0.2133 (for 37 degrees of freedom). Restricting to each redshift bin alone, the $\Delta\chi^2$ to the best-fit is $\left(4.2,  -0.3,  4.3,  5.9\right)$ for $i=\left(1,2,3,4\right)$.

The best-fit value of $A$ is shown, along with the corresponding distribution from the null simulations, on the right hand side of Figure~\ref{fig:chi2sdistribution_auto_fitfacs}. We overplot a Gaussian distribution with mean 0 and standard deviation given by the measured standard deviation of the values recovered from the simulations; this can be interpreted as the $1\sigma$ errorbar on $A$. We recover a best-fit value of $A=0.638\pm0.17$, providing $3.8\sigma$ evidence of $A\ne0$. We note that again only 2 simulations out of 6800 have a best fit value of $A$ with $|A|$ greater than the measured value.

The datapoints, along with the  best-fit model, are shown in Figure~\ref{fig:bestfit_and_Data}.  For visualization purposes, it is also helpful to co-add the datapoints from the different redshift bins and compare to a co-added best-fit model. 

In order to co-add the datapoints, we first divide each measurement (and the covariance) by the relevant best-fit theory, and multiply by a common reference signal. For the reference signal, we take the geometric mean of the best-fit theory:
\begin{align}
A_{\mathrm{best-fit}}C_L^{VV}= \left(\prod_{i=I=1}^4 A_{\mathrm{best-fit}}C_L^{v^{i}v^{I}}\right)^{1/4}.
\end{align}
Thus our rescaled measurements are given by
\begin{align}
\hat d^i_L = \frac{A_{\mathrm{best-fit}}C_L^{VV}}{A_{\mathrm{best-fit}}C_{L}^{v^{i}v^{J}}}\hat C_L ^{\hat v_r^{\mathrm{kSZ},i}\hat v_r^{\mathrm{template},I}}.
\end{align}
These are estimators of $A_{\mathrm{best-fit}}C_L^{VV}$. We then optimally co-add the datapoints given the covariances,
\begin{equation}
\hat d^{\mathrm{co-added}}_L= \sum _i w^i_L \hat d^i_L,
\end{equation}
where the weights $w_i$ are such that the variance of $\hat d_L^{\mathrm{co-added}}$ is minimized subject to the constraint that $\sum_i w^i_L=1$. This is the minimum-variance estimator of $A_{\mathrm{best-fit}}C_L^{VV}$. The weights are found straightforwardly by evaluating
\begin{align}
w^i_L = \frac{a^T \mathbb C^{-1}_L}{a^T \mathbb C^{-1}_L a^T},
\end{align}
where $\mathbb C_L^{-1}$ is the inter-bin inverse of the covariance matrix $\mathbb C$, and $a$ is a vector of ones. We show the optimally co-added datapoints, and the reference signal, in Figure~\ref{fig:reweighted_theory}.

\subsection{Scale-dependence of the signal-to-noise}\label{sec:scaledep}

The measured best-fit value of $A$, as a function of the minimum multipole included in the analysis $L_{\mathrm{min}}$, is shown in Figure~\ref{fig:minmultipole}. As expected, much of the signal-to-noise is on the largest scale, with a $3\sigma$ signal requiring $L_{\mathrm{min}}\le12$, and a $2\sigma$ signal requiring $L_{\mathrm{min}}<20$. This indicates the constraining power of the extremely large scales, and simultaneously the stability of the measurement to the value of $L_{\mathrm{min}}$.

\subsection{Redshift-dependence of the signal}\label{sec:redshift_dependence}

We have previously been fitting one best-fit amplitude jointly to all redshift bins. We can also allow for redshift dependence of the signal by fitting a different parameter $A^i$ separately each redshift bin. We show the results of such a procedure in Figure~\ref{fig:redshift_dependence}. On the left of Figure~\ref{fig:redshift_dependence} we show Gaussian distributions centred on the measured best-fit value of $A_i$, with standard deviations given by the measured standard deviation of the best-fit values on the 6800 simulations. We also show the same quantity for our jointly-fit $A$. Note that all of the measurements are consistent with each other, and so can be combined. On the right, we make any possible redshift dependence clear. There is not enough signal to noise to detect any trend. The $\chi^2$ with respect to the four-parameter model is 41.90, which corresponds to a PTE of 13.8\% (for 33 degrees of freedom). Recalling that the fixed-$A$ model had a $\chi^2$ of 43.53 (and a PTE of 0.2133), it is clear that we cannot prefer the varying-$A$ model over the fixed-$A$ model---ie, we do not have the signal-to-noise to detect evolution in $A$.

\section{Conclusions}

In this work, we have found the first evidence, at $3.8\sigma$, for the large-scale velocity field as reconstructed by kSZ tomography using the 2-dimensional ``lightcone'' formalism. This formalism (which is used to reconstruct and model the 2-dimensional \textit{angular} velocity field) stands in contrast to the 3-dimensional formalism, which is more appropriate for kSZ velocity reconstruction with spectroscopic surveys in which full 3-dimensional information is available. 

Currently, this measurement serves as a proof-of-principle that such a signal is detectable. However, given the cosmological interest in this observable, and the impending increase  in galaxy samples and improvements in small-scale CMB data, we expect kSZ tomography to become a precision probe of the late-Universe density field in the near-term future. Such a research programme will require well-tested reconstruction pipelines, appropriate for various datasets.

The 2-dimensional lightcone formalism we have used is complementary to the 3-dimensional ``box'' formalism (e.g., as developed in~\cite{2018arXiv181013423S}). The box formalism makes it easy to incorporate the full 3-dimensional information in a spectroscopic survey; however it has drawbacks in that it is difficult to incorporate redshift evolution and very large-scale curved-sky effects. The 2-dimensional lightcone formalism, in contrast, naturally incorporates these effects; but the requirement of tomographic redshift binning necessarily loses some of the 3-dimensional information (which can dilute the velocity signal). For photometric surveys, however, there will be no information loss if the redshift bins are chosen to have widths similar to the photometric redshift error of the sample.  As such, it will be important to develop both pipelines in preparation for different types of galaxy samples. 

This work was intended as a demonstration and proof-of-principle of the kSZ velocity reconstruction pipeline on data, and so we report only our detection significance and best-fit value of the optical depth bias $A$. However, we note that the $C_L^{vv}$ signal we measure  is directly sensitive to interesting beyond-$\Lambda$CDM physics including local primordial non-Gaussianity, and can be used directly to constrain $f_{\mathrm{NL}}$; we leave such a constraint for follow-up work. {Indeed, higher signal-to-noise may be accessible using current datasets with velocity reconstruction performed directly on the photometric galaxies, as advocated in~\cite{2024PhRvD.109j3533R,2024PhRvD.109j3534H} (and applied recently to a similar combination of datasets in~\citealt{2024arXiv240707152H})}.  We note that~\cite{2024arXiv240805264K} found constraints on $f_{\rm NL}$ with $\sigma (f_{\rm NL})\sim200$ from applying this method to the \textit{Planck}+unWISE data combination.

We note that the galaxy sample we use has previously been used to constrain $f_{\mathrm{NL}}$~\citep{2024MNRAS.532.1902R}; in principle, the combination with the kSZ velocity observable 
 can allow for improvements compared to the auto power spectrum measurement---both by way of sample variance cancellation~\citep{2009PhRvL.102b1302S} and because the $C_L^{v^{\mathrm{kSZ}}v^{\mathrm{template}}}$ constraint \textit{alone} is interesting due to its dependence on different large-scale systematics compared to the auto power spectrum. As was demonstrated on data for the first time recently in~\cite{2024arXiv240805264K}, there is also a wide range of beyond-$\Lambda$CDM models that can be constrained with the kSZ velocity signal.

\section*{Acknowledgements}

Support for ACT was through the U.S.~National Science Foundation through awards AST-0408698, AST-0965625, and AST-1440226 for the ACT project, as well as awards PHY-0355328, PHY-0855887 and PHY-1214379. Funding was also provided by Princeton University, the University of Pennsylvania, and a Canada Foundation for Innovation (CFI) award to UBC. ACT operated in the Parque Astron\'omico Atacama in northern Chile under the auspices of the Agencia Nacional de Investigaci\'on y Desarrollo (ANID). The development of multichroic detectors and lenses was supported by NASA grants NNX13AE56G and NNX14AB58G. Detector research at NIST was supported by the NIST Innovations in Measurement Science program. Computing for ACT was performed using the Princeton Research Computing resources at Princeton University, the National Energy Research Scientific Computing Center (NERSC), and the Niagara supercomputer at the SciNet HPC Consortium. SciNet is funded by the CFI under the auspices of Compute Canada, the Government of Ontario, the Ontario Research Fund–Research Excellence, and the University of Toronto. We thank the Republic of Chile for hosting ACT in the northern Atacama, and the local indigenous Licanantay communities whom we follow in observing and learning from the night sky.

FMcC acknowledges support from the European Research Council (ERC) under the European Union's Horizon 2020 research and innovation programme (Grant agreement No.~851274). 
The Flatiron Institute is a division of the Simons Foundation. EC acknowledges support from the European Research Council (ERC) under the European Union’s Horizon 2020 research and innovation programme (Grant agreement No.~849169). MCJ is supported by the National Science and Engineering Research Council through a Discovery grant. This research was supported in part by Perimeter Institute for Theoretical Physics. Research at Perimeter Institute is supported by the Government of Canada through the Department of Innovation, Science and Economic Development Canada and by the Province of Ontario through the Ministry of Colleges and Universities. KM acknowledges support from the National Research Foundation of South Africa. CS acknowledges support from the Agencia Nacional de Investigaci\'on y Desarrollo (ANID) through Basal project FB210003. This work was supported by a grant from the Simons 
Foundation (CCA 918271, PBL).  We thank Boryana Hadzhiyska for useful discussions.

This work made use of HEALPix~\citep{2005ApJ...622..759G} and its python implementation \texttt{healpy}~\citep{Zonca2019}.\footnote{\url{https://healpy.readthedocs.io/en/latest/}}
%%%%%%%%%%%%%%%%%%%%%%%%%%%%%%%%%%%%%%%%%%%%%%%%%%

%%%%%%%%%%%%%%%%%%%% REFERENCES %%%%%%%%%%%%%%%%%%

%\bibliographystyle{mnras}
\bibliographystyle{act_titles}
\bibliography{references}

%%%%%%%%%%%%%%%%%%%%%%%%%%%%%%%%%%%%%%%%%%%%%%%%%%

%%%%%%%%%%%%%%%%% APPENDICES %%%%%%%%%%%%%%%%%%%%%

\appendix

\section{Galaxy-electron cross correlation}\label{app:cltaug}

We model the galaxy-electron cross correlation $C_\ell^{g\tau}$ using a halo model. Within the halo model, all dark matter is placed in discrete \textit{halos}, which we model as spherically-symmetric; the lowest-density regions of the dark matter field, which are outside of the halos, are modelled as having zero dark matter. Cosmological observables are simplistically modelled as depending only on the mass and redshift of the halo; this ignores several other real effects on the observables due to other factors such as the environment and merger history of the halo.

For a review of the halo model, see~\cite{2002PhR...372....1C}. We perform all halo model calculations with \texttt{class\_sz}~\citep{2023arXiv231018482B,2023JCAP...03..039B}; those references describe in detail the calculations we use. \texttt{class\_sz} is based on the Boltzmann code \texttt{class}~\citep{2011arXiv1104.2932L,2011JCAP...07..034B}.\footnote{\url{http://class-code.net}} For completeness, we describe them briefly here (including listing our specific parameter and modelling choices, where relevant).

\subsection{2-dimensional (angular) power spectra: $C_\ell^{g\tau}$ and 3-dimensional power spectrum $P_{ge}(k,z)$}

In general, $C_\ell^{g\tau}$ comprises a 2-halo term quantifying inter-halo correlations (ie clustering) and a 1-halo term quantifying intra-halo correlations (which, as such, is sensitive to the radial profile of halos).

Within the Limber approximation, $C_\ell^{g\tau}$ is a weighted integral over redshift of the three-dimensional power spectrum
\begin{equation}
C_\ell^{g^i\tau}=\int \frac{d\chi}{\chi^2} W^i(\chi)W^\tau(\chi)\left(P_{ge}\left(k=\frac{\ell}{\chi},z\right) \right),
\end{equation}
where  $W^i(\chi)$ is the galaxy window function, $W^\tau(\chi)$ is the electron window function, and $P_{ge}(k,z)$ is the three-dimensional electron-galaxy power spectrum. The galaxy window functions we use are 
\begin{equation}
W^i(\chi) =\frac{\frac{dN^i}{d\chi}}{\int d\chi \frac{dN^i}{d\chi}}.
\end{equation}
The electron window function is
\begin{equation}
W^\tau(\chi) = \frac{\sigma_T}{1+z}n_e(\chi) ,
\end{equation}
where $n_e(\chi)$ is the average number density of electrons at $\chi$
\begin{align}
n_e (\chi) = \frac{\rho_e}{m_p \mu_e},
\end{align}
with $\rho_e$ the electron density, $m_p$ the mass of the proton, and $\mu_e\sim1.14$ the mean molecular weight per electron (thus $m_p \mu_e$ is the mean molecular mass per electron). $\rho_e$ is modelled as
\begin{align}
\rho_e(\chi) = f_{\mathrm{free}}\rho_b = f_{\mathrm{free}}f_b\Omega_m (1+z)^3\rho_{\mathrm{crit}}
\end{align}
where $f_{\mathrm{free}}$ is the fraction of free electrons (we assume $f_{\mathrm{free}} =1$) and $\rho_b=f_b \Omega_m (1+z)^3 \rho_{\mathrm{crit}}$ is the mean baryon density at $z$, with $f_b=\frac{\Omega_b}{\Omega_m}$ the baryon fraction, and $\rho_{\mathrm{crit}}$ the critical density today.

The three-dimensional galaxy-electron power spectrum comprises the 2-halo and 1-halo information according to
\begin{align} 
P_{ge}(k,z) = 
P_{ge}^{2h}(k,z) +
P_{ge}^{1h}(k,z) 
\end{align}
where the 2-halo term is given by
\begin{align}
P_{ge}^{2h}(k,z) = &\left(\int dMb(M,z)\frac{dN}{dM}u_g(k,M,z)\right)\nonumber\\
&\times\left(\int dMb(M,z)\frac{dN}{dM} \frac{M}{\rho_m}u_e(k,M,z)\right)
\end{align}
and the 1-halo term is given by
\begin{align}
P_{ge}^{1h}(k,z) = \int dM \left(\frac{M}{\rho_m}u_e(k,M,z)u_g(k,M,z)\right).
\end{align}
Above, $M$ indicates halo mass,   $b(M,z)$ is the halo bias, and $\frac{dN}{dM}$ (which is also a function of $(M,z)$)  is the halo mass function.  $u_e(k,M,z)$ is the Fourier-transform of the spherically symmetric electron profile, and $u_g(k,M,z)$ is the Fourier-transform of the spherically symmetric galaxy profile.

We integrate over the mass range $1\times10^{10}h^{-1}M_\odot<M<5\times10^{15}h^{-1}M_\odot$. We use the halo mass function of~\cite{2008ApJ...688..709T} and the halo bias of~\cite{2010ApJ...724..878T}, and the halo mass-concentration relation of~\cite{2013ApJ...766...32B}.

\subsection{Galaxy model}
To model the galaxy distribution, we use a halo occupation distribution (HOD) model:
\begin{align}
u_g(k,M,z) = \frac{N_c(M)+N_s(M)u(k,M,z)}{n_g},
\end{align}
where $n_g$ is the mean galaxy number density, $N_c(M)$ is the expected number of central galaxies, $N_s(M)$ is the expected number of satellited galaxies, and the satellites are distributed according to $u(k,M,z)$ which we take to be (the Fourier transform of) a Navarro--Frenk--White (NFW) profile~\citep{1997ApJ...490..493N}. The mean galaxy density $n_g$ can be calculated by integrating over $N_c$ and $N_s$:
\begin{align}
n_g = \int dM\frac{dN}{dM}\left(N_c(M)+N_s(M)\right).
\end{align}

The mean numbers of centrals $N_c(M)$ and of satellites $N_s(M)$ take the forms
\begin{align}
N_c(M) =& \frac{1}{2}\left(1+\mathrm{erf}\left(\frac{\log M - \log M^{\mathrm{HOD}}_{\mathrm{min}}}{\sigma_{{\log M}}}\right)\right),\\
N_s(M) =& N_c(M)\left(\frac{M-M_0}{M_1^\prime}\right)^{\alpha_s}.
\end{align}
A similar HOD model has been previously fit to the DESI galaxies~\citep{2024MNRAS.530..947Y}. We choose parameters which are similar (although not exactly the same) as the best-fit parameters from this fit; we list them in Table~\ref{tab:HOD_params}.

\begin{table}
\centering
\begin{tabular}{|c|c|}\hline
Parameter&Value\\\hline
$M_{\mathrm{min}}^{\mathrm{HOD}}$ & $10^{14.08} h^{-1} M_\odot$\\
$\sigma_{\log M}$ & 0.27\\
$M_0$ & $0.65\times10^{12.89}h^{-1} M_\odot$\\
$\alpha_s$ & 1.2\\
$M_1^\prime$ & $10^{14.8} h^{-1} M_\odot$\\\hline
\end{tabular}
\caption{The parameters used in the HOD to describe the DESI galaxies.}\label{tab:HOD_params}
\end{table}

\subsection{Electron model}

For the electrons, we use the ``AGN-feedback'' generalized NFW (gNFW) profiles of~\cite{2016JCAP...08..058B}.
%%%%%%%%%%%%%%%%%%%%%%%%%%%%%%%%%%%%%%%%%%%%%%%%%%

\section{Filters for the quadratic estimator}\label{app:filters}

In this Appendix we plot the filters that we use for the galaxy and temperature fields in the quadratic estimator.

The quadratic estimator for the velocity is 
\begin{equation}
\hat v_r^{i}{}_{LM} = N_L^i \left(\tilde T(\hat n) \zeta^{i}(\hat n)\right)_{LM}
\end{equation}
where
\begin{align}
\zeta_{lm}^{i}=& C_\ell^{\tau g^i}\tilde{\delta ^{g^i}}_{\ell m},
\end{align}
with the tilde indicating an inverse-variance filtered field
\begin{align}
\tilde X_{\ell m} \equiv \frac{X_{\ell m}}{C_\ell^{XX}}.
\end{align}
The variance filter $C_\ell^{XX}$ includes all sources of signal and noise in $X$.

\subsection{Temperature filter}
\begin{figure}
\includegraphics[width=\columnwidth]{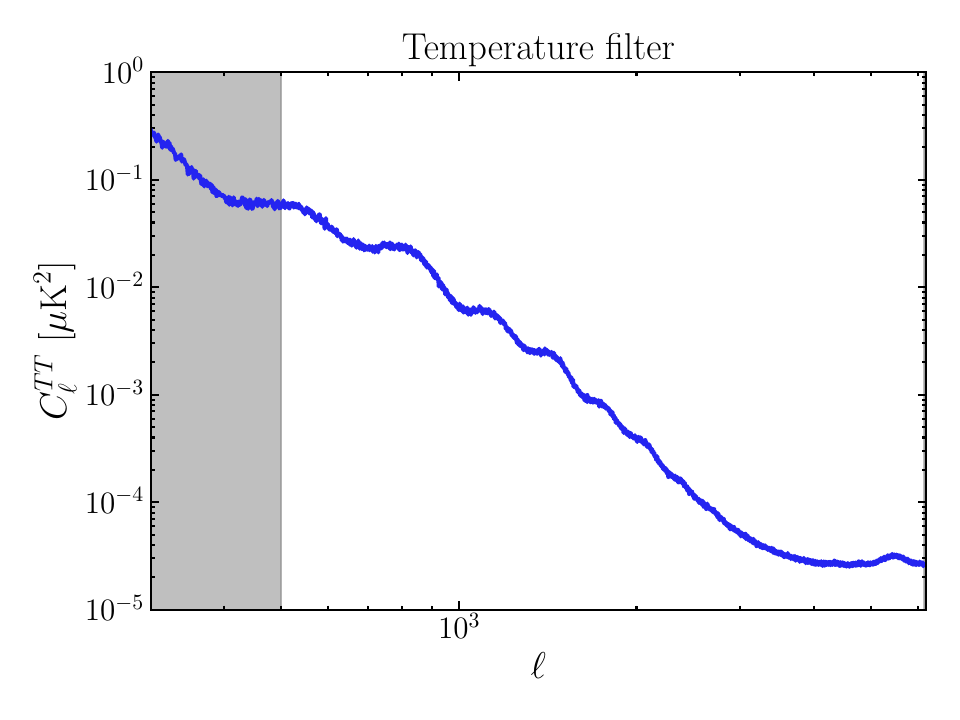}
\caption{The filter $C_\ell^{TT}$ by which the temperature field is inverse-variance filtered. This is measured directly from the pseudo-$C_\ell$s of the field using \texttt{healpy.anafast()} (and dividing by the sky area $f_{\mathrm{sky}}$). We do not use any of the multiples at $\ell<500$ (indicated by the gray region). }\label{fig:tfilter}
\end{figure}

The temperature filter $C_\ell^{TT}$ used to filter the temperature field $T(\hat n)$ to create $\tilde T(\hat n)$ is estimated directly from the temperature field by calculating the power spectrum using \texttt{hp.anafast()} on the region left unmasked by the ACT mask. This is shown in Figure~\ref{fig:tfilter}. We have checked that are results are insensitive to whether we bin and smooth the measured power spectra using large multipole bins, or we use the raw $C_\ell$s.

\subsection{Galaxy filters}

\begin{figure}
\includegraphics[width=\columnwidth]{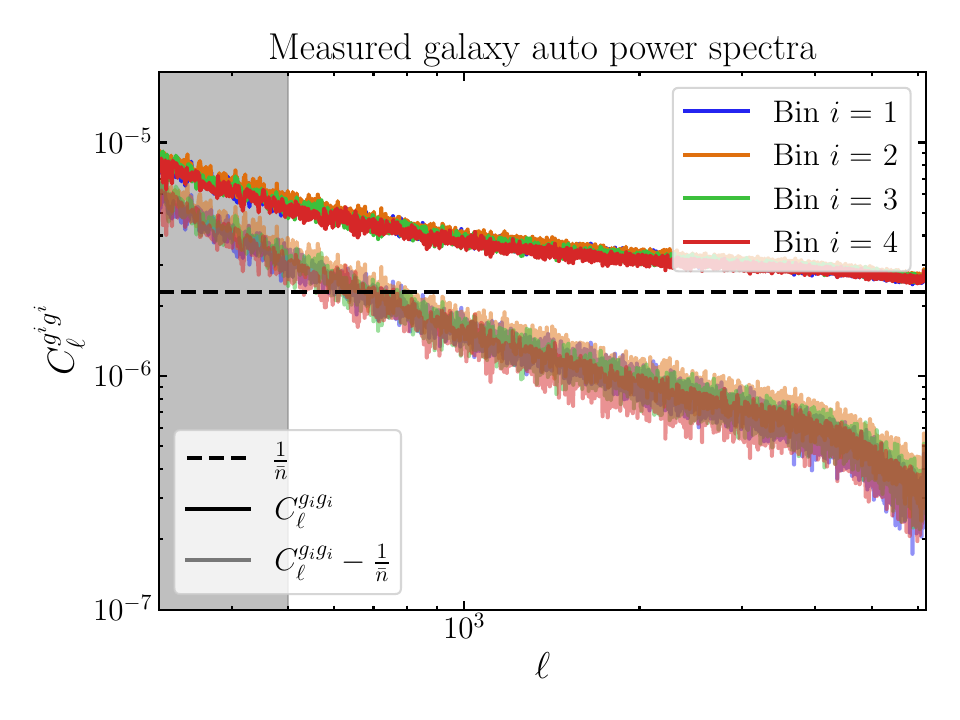}
\caption{Measured galaxy auto power spectra for the DESI LRGs, estimated directly from the maps with \texttt{healpy.anafast()} (and division by $f_{\mathrm{sky}}$). We also indicate the Poissonian shot noise level $\frac{1}{\bar n}$, where $\bar n$ is the mean number density of the galaxies (recall that the samples are chosen such that each has an equal number of objects $N=2409741$, so that this shot noise is equal for all bins). Additionally we indicate the measured auto power spectrum with this shot noise subtracted (although note that this quantity is never used and is only included here for illustrative purposes). }\label{fig:galaxyautopower}
\end{figure}

We use four photometrically binned galaxy samples such that $I={1,2,3,4}$.  The galaxy filters used in each estimation to create $\zeta^{i}(\hat n)$ are thus $\frac{C_\ell^{\tau g^i}}{C_\ell^{g^ig^i}}$. Again, we estimate $C_\ell^{g^ig^i}$ by directly measuring the pseudo-$C_\ell$s of the overdensity maps, on the region left unmasked by the DESI LRG masks. These are shown in Figure~\ref{fig:galaxyautopower}.

\begin{figure}
\includegraphics[width=\columnwidth]{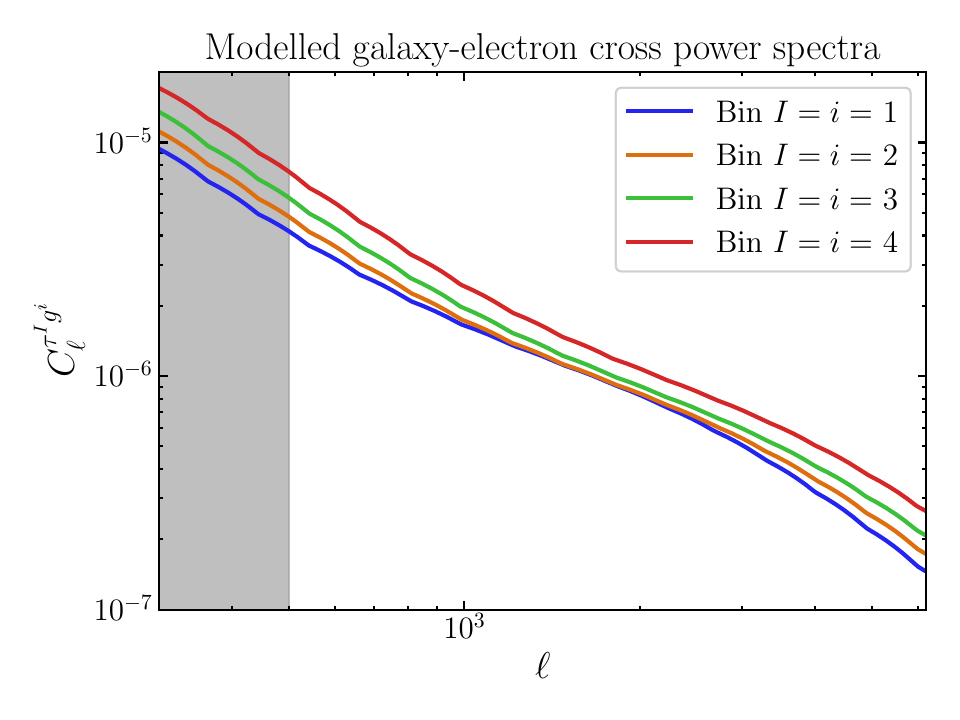}
\caption{Modelled galaxy-electron cross power spectra, modelled using \texttt{class\_sz} as described in Appendix~\ref{app:cltaug}. }\label{fig:cltaug_plot}
\end{figure}

\begin{figure}
\includegraphics[width=\columnwidth]{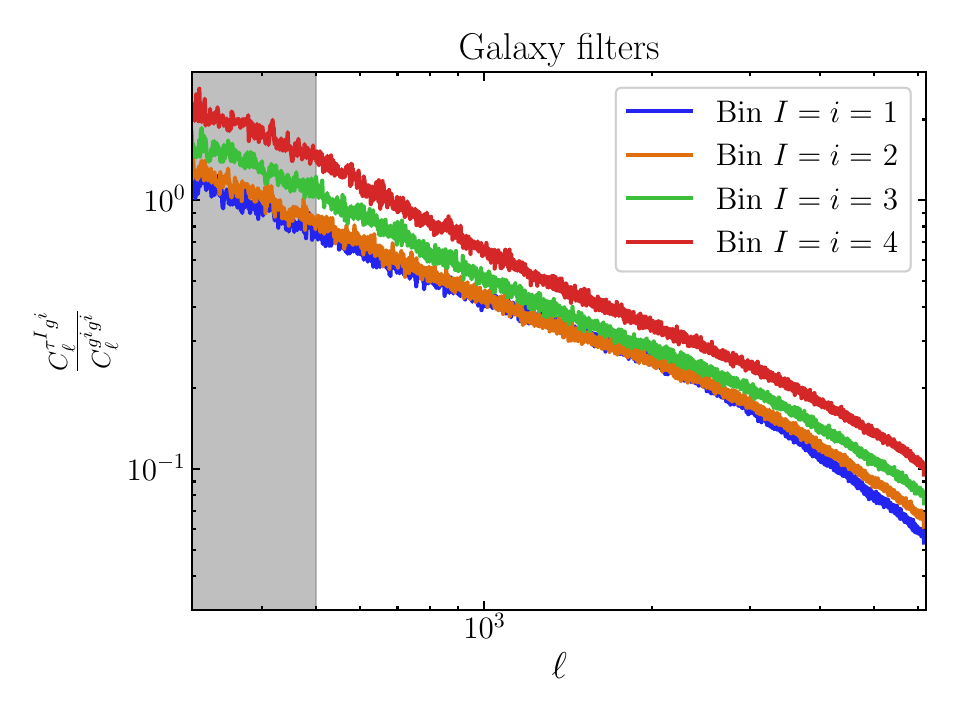}
\caption{The harmonic filters used to filter the $\delta ^{g^i}$ fields to create $\zeta^{i}$, $\frac{C_\ell^{\tau g^i}}{C_\ell^{g^i g^i}}$.}\label{fig:galaxy_filters}
\end{figure}

The $C_\ell^{\tau g^i}$ are  modelled with \texttt{class\_sz} as described in Appendix~\ref{app:cltaug}. The models are shown in Figure~\ref{fig:cltaug_plot}. The combined filters $\frac{C_\ell^{\tau g^i}}{C_\ell^{g^ig^i}}$ are then shown in Figure~\ref{fig:galaxy_filters}.  Again, we have checked that are results are insensitive to whether we bin and smooth the measured power spectra using large multipole bins, or we use the raw $C_\ell$s.

\subsection{Galaxy auto power spectra}

\begin{figure*}
    \includegraphics[width=0.49\textwidth]{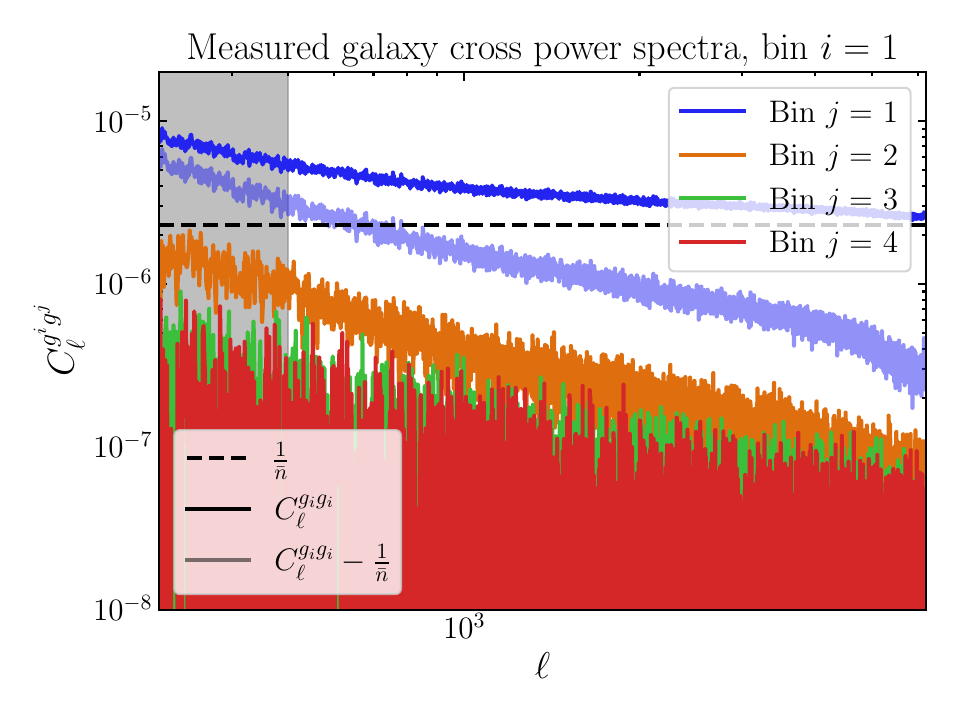}
    \includegraphics[width=0.49\textwidth]{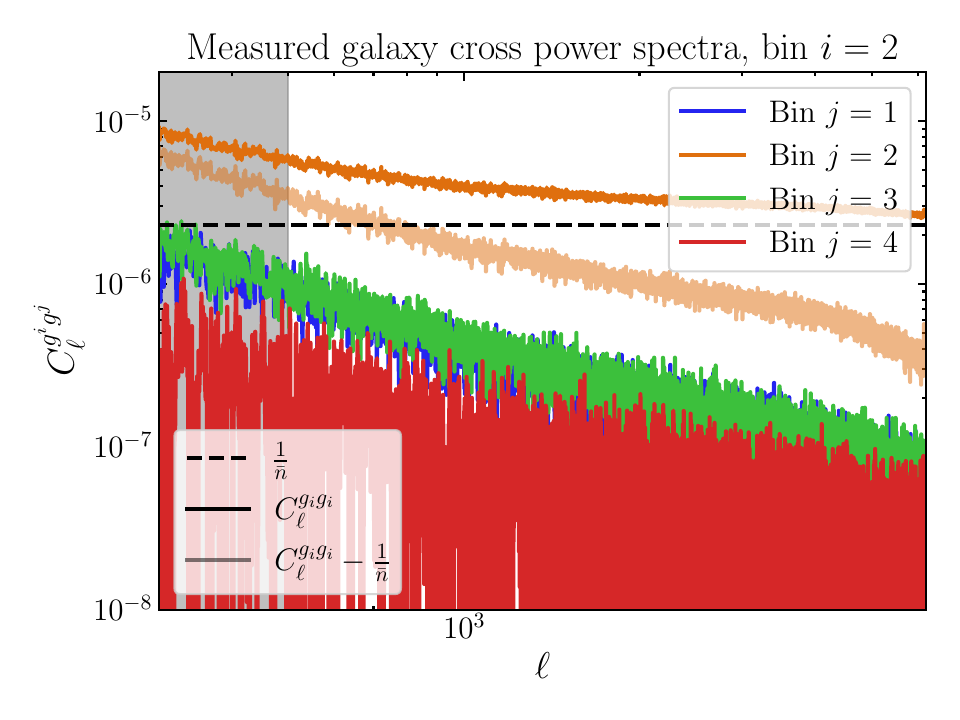}
    \includegraphics[width=0.49\textwidth]{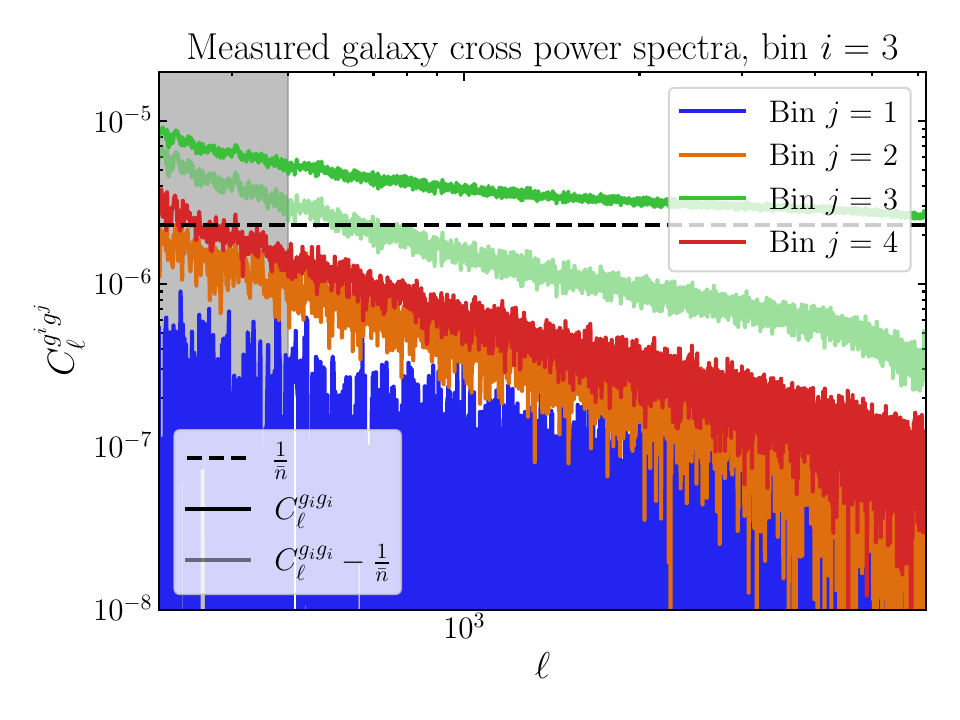}
    \includegraphics[width=0.49\textwidth]{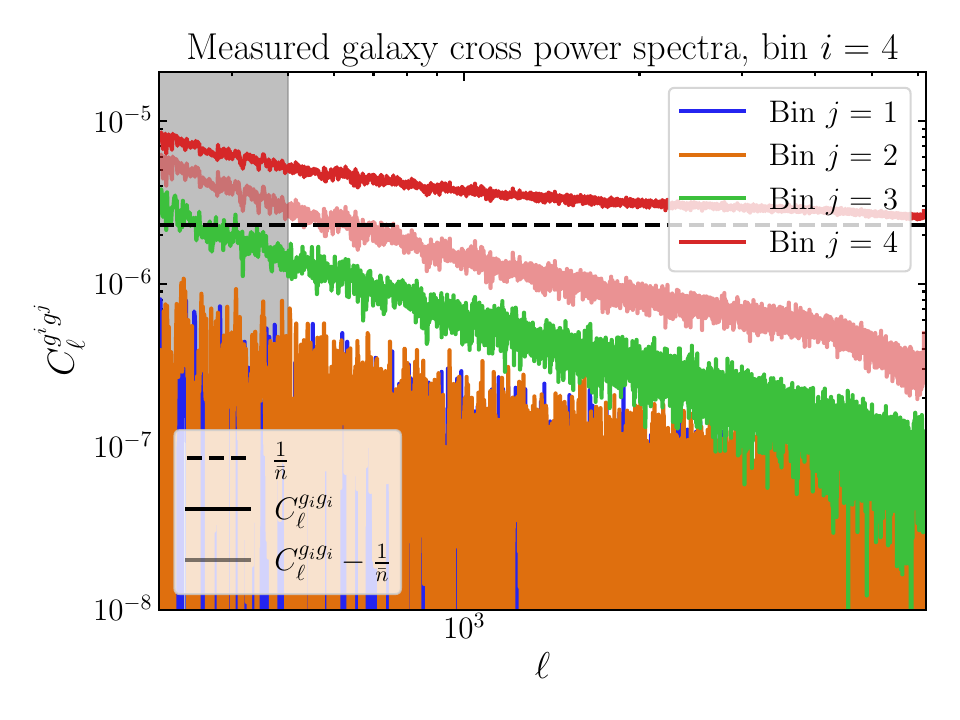}

    \caption{The cross power spectra between the different galaxy bins. Note that for bins that are not neighbouring, the measurement is not significantly non-zero (as expected, given the low overlap between the redshift bins and the small scales involved---in the Limber approximation, a model would predict exactly zero correlation). {In each plot, we also show the power spectra with an estimate of the shot noise subtracted (with the estimate calculated simply as $\frac{1}{\bar n}$).}}
    \label{fig:galaxycrosspower}
\end{figure*}

To correctly account for the covariance between the measured $C_L^{v^i v^I}$ and $C_L^{v^i v^J}$ the covariance between the different galaxy samples ($i\ne j$)  must be included. In practice we do this by creating a covariance matrix using simulations that display the appropriate covariance. In particular, we draw Gaussian simulations with the correct covariance matrix $C_\ell^{g^ig^j}$. The auto spectra $C_\ell^{g^ig^i}$ were shown in Figure~\ref{fig:galaxyautopower}; we show the cross power spectra in Figure~\ref{fig:galaxycrosspower}. These are similarly estimated by measuring the cross- pseudo $C_\ell$s of the two maps $\delta ^{g^i}$ and $\delta ^{g^j}$  using \texttt{healpy.anafast()} and dividing by $f_{\mathrm{sky}}$. Note that the cross-spectra are only non-zero for neighbouring bins, as expected given the small overlap of their redshift kernels.

\section{Velocity normalization}\label{app:velocity_normalization}

\begin{figure*}
\includegraphics[width=0.49\textwidth]{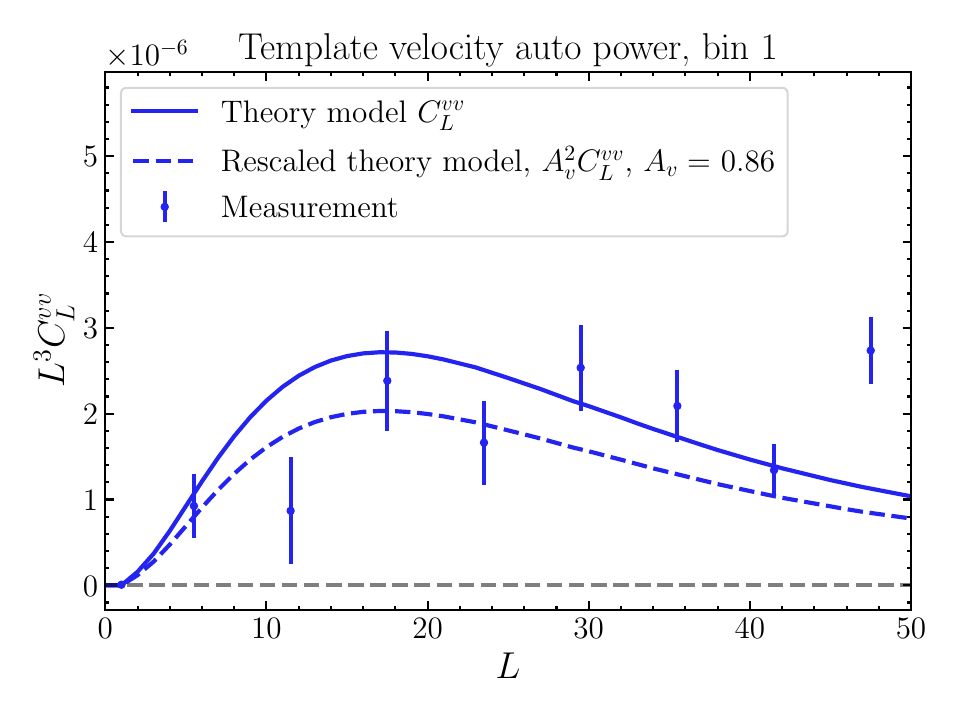}
\includegraphics[width=0.49\textwidth]{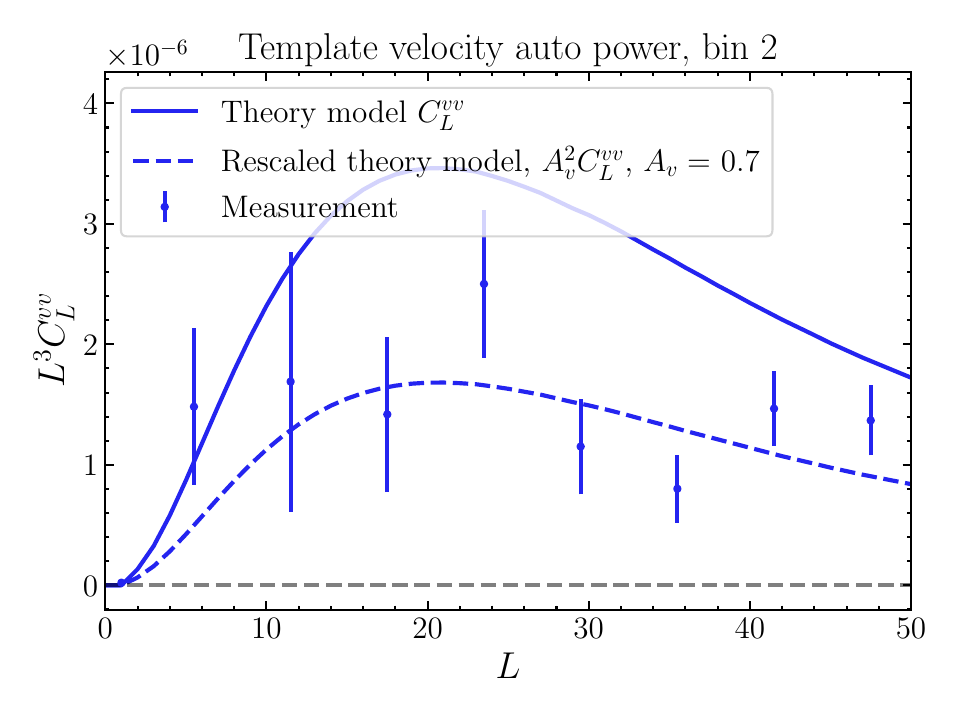}
\includegraphics[width=0.49\textwidth]{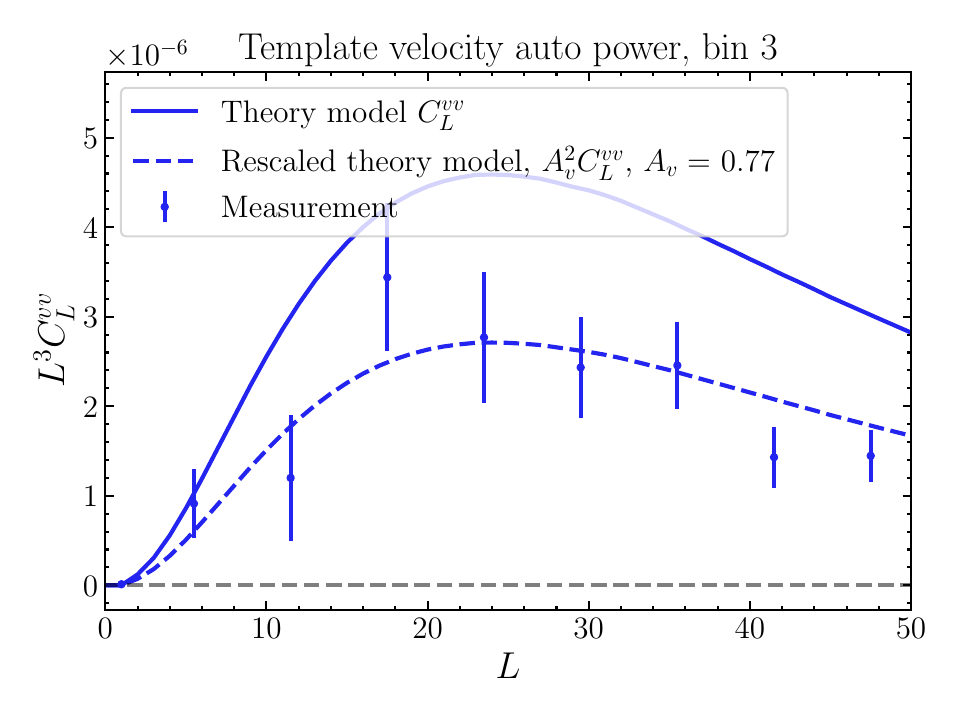}
\includegraphics[width=0.49\textwidth]{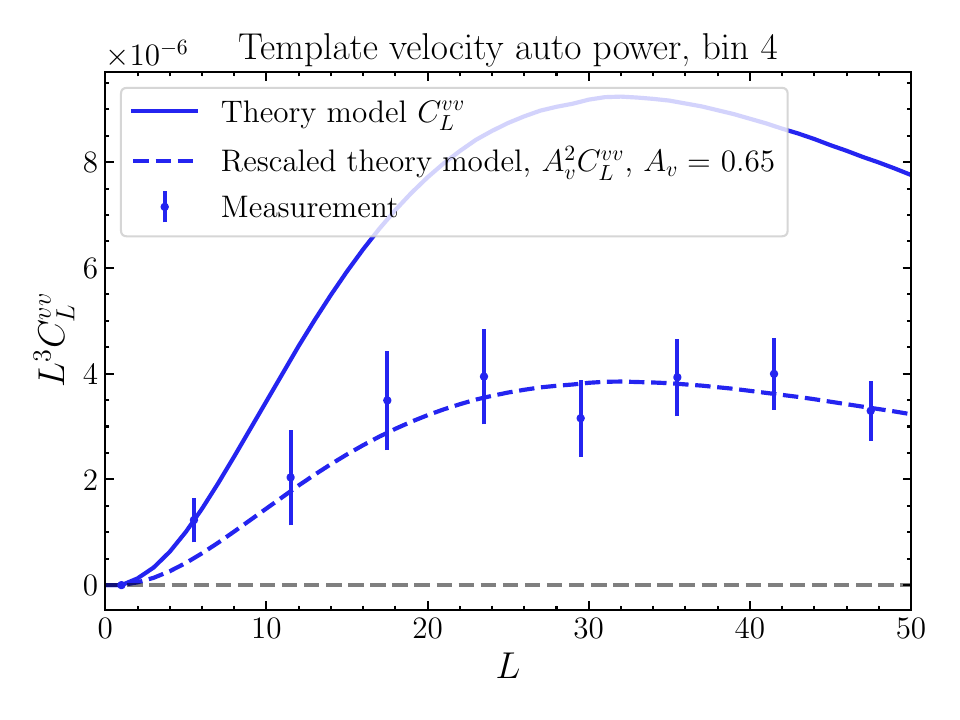}
\caption{The measured autospectra of the velocity templates, along with the $\Lambda$CDM model predictions (solid) and the models rescaled by a normalization {to empirically correct for the impact of Wiener filtering in generating the templates,} such that they better describe the data. }\label{fig:autospectra_template}
\end{figure*}

The continuity-equation velocities have been reconstructed from SDSS with a Wiener filter applied, and their normalization should be inferred using Monte-Carlo methods. Alternatively, we compare the power spectra of each of our templates to those expected from a $\Lambda$CDM calculation, and normalize by the appropriate amount to correct this. The measured power spectra are shown  in Figure~\ref{fig:autospectra_template}.

\section{Reconstructed velocity maps}\label{app:reconmaps}
\begin{figure*}
\includegraphics[width=0.49\textwidth]{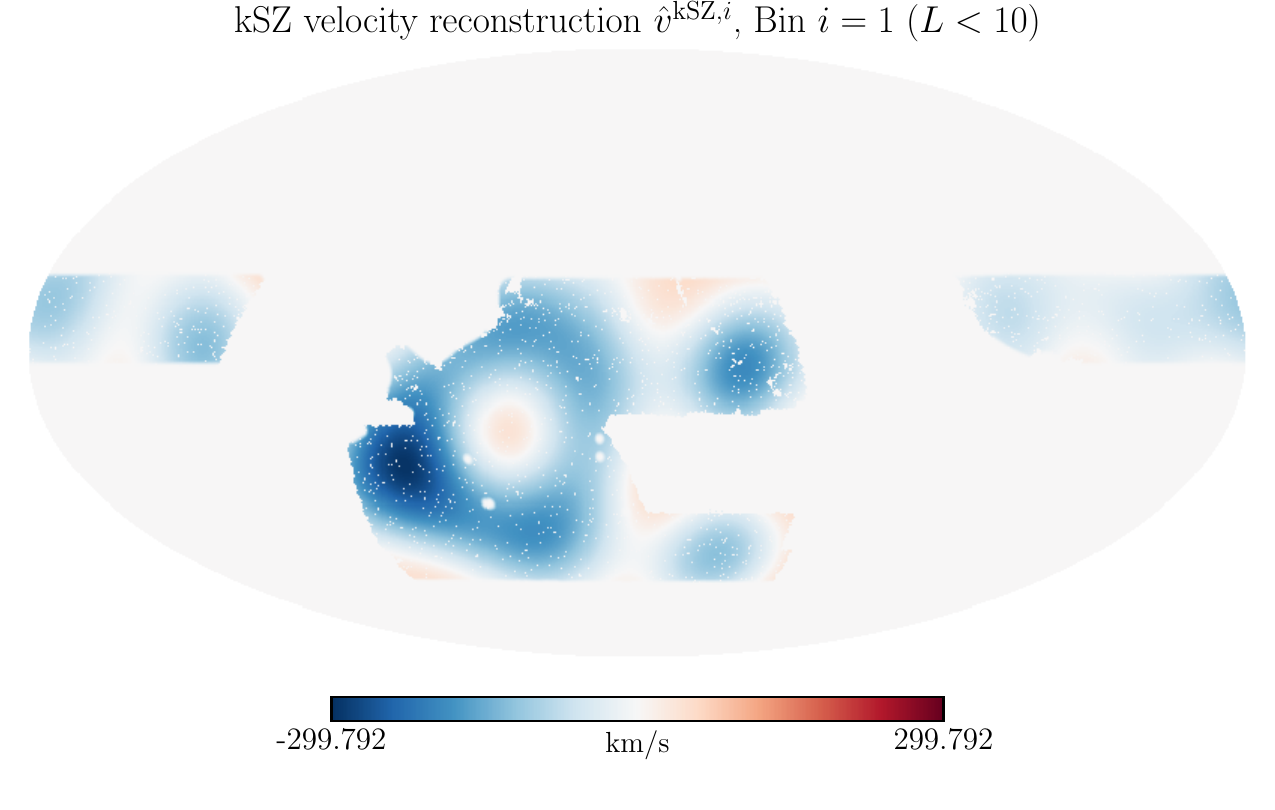}
\includegraphics[width=0.49\textwidth]{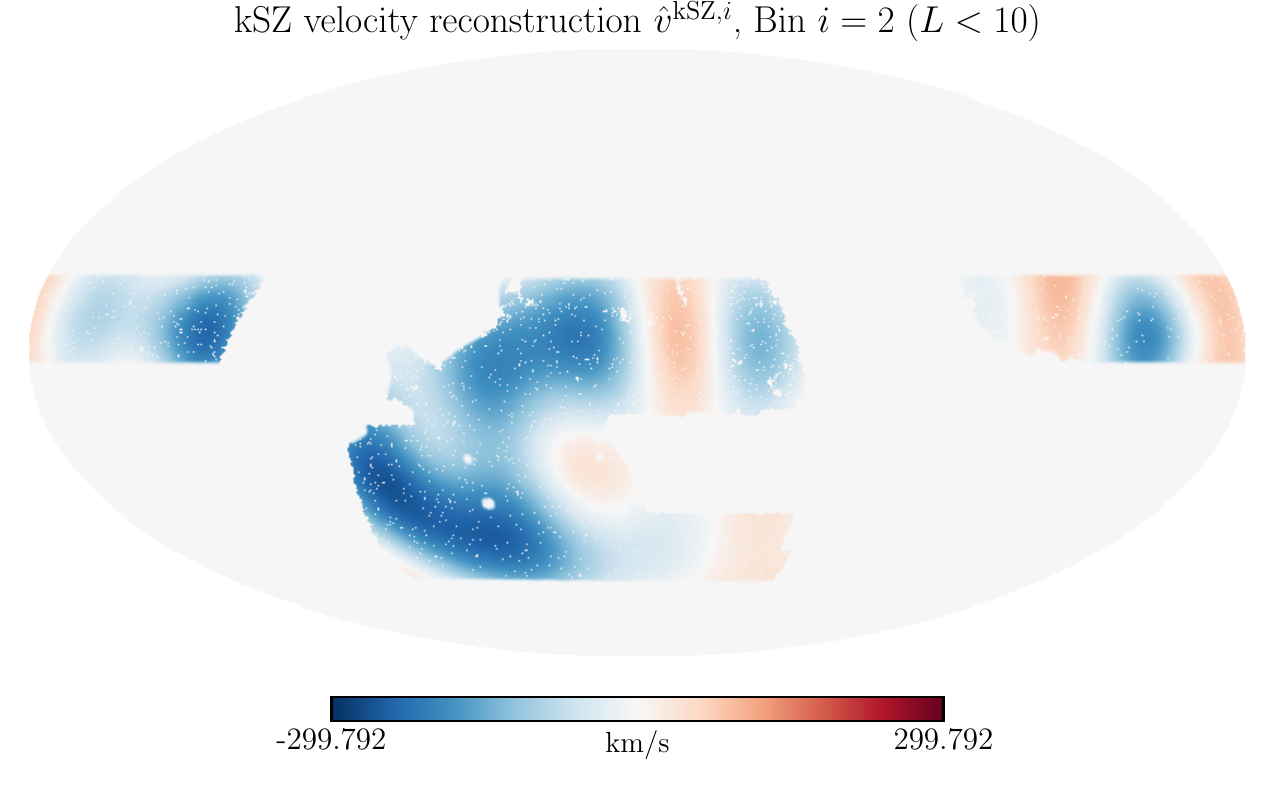}
\includegraphics[width=0.49\textwidth]{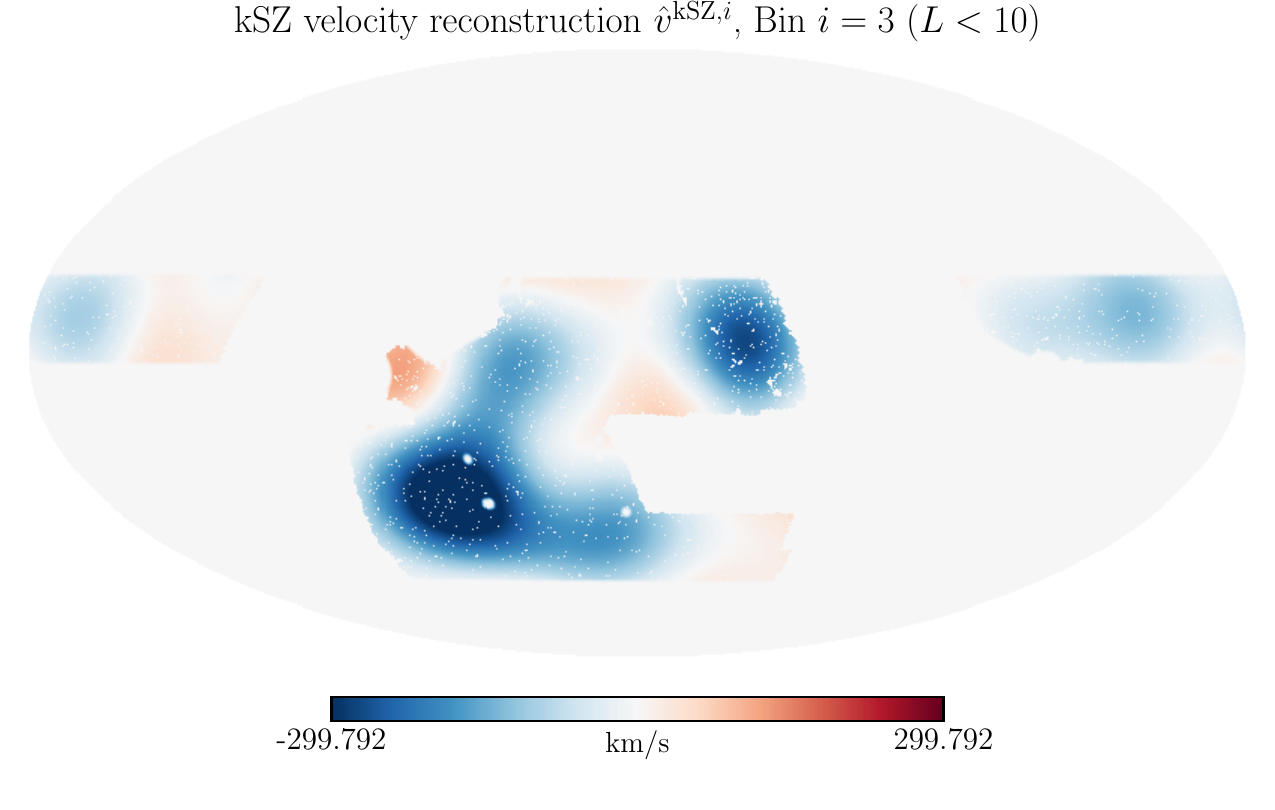}
\includegraphics[width=0.49\textwidth]{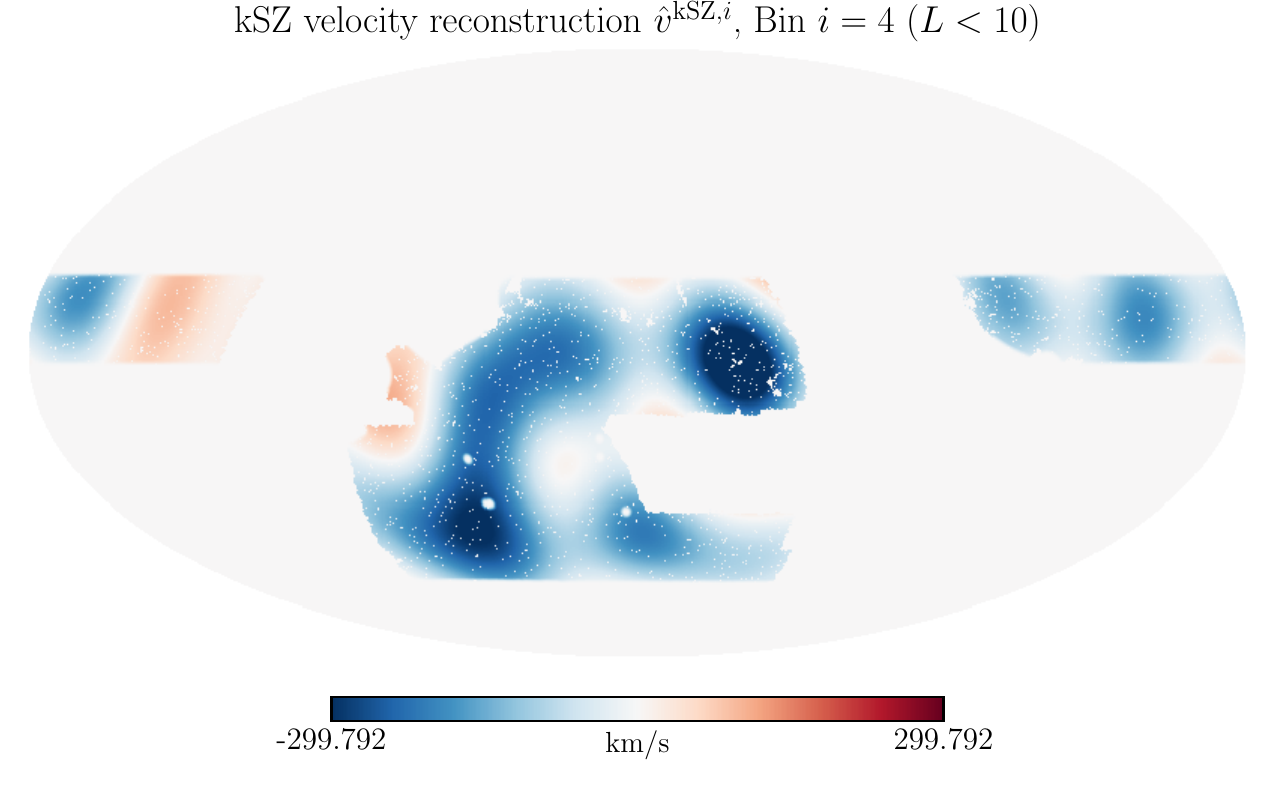}
\caption{The reconstructed velocity, filtered to preserve only $L<10$ information.}\label{fig:velocity_reconstructed_maps_l10}

\includegraphics[width=0.49\textwidth]{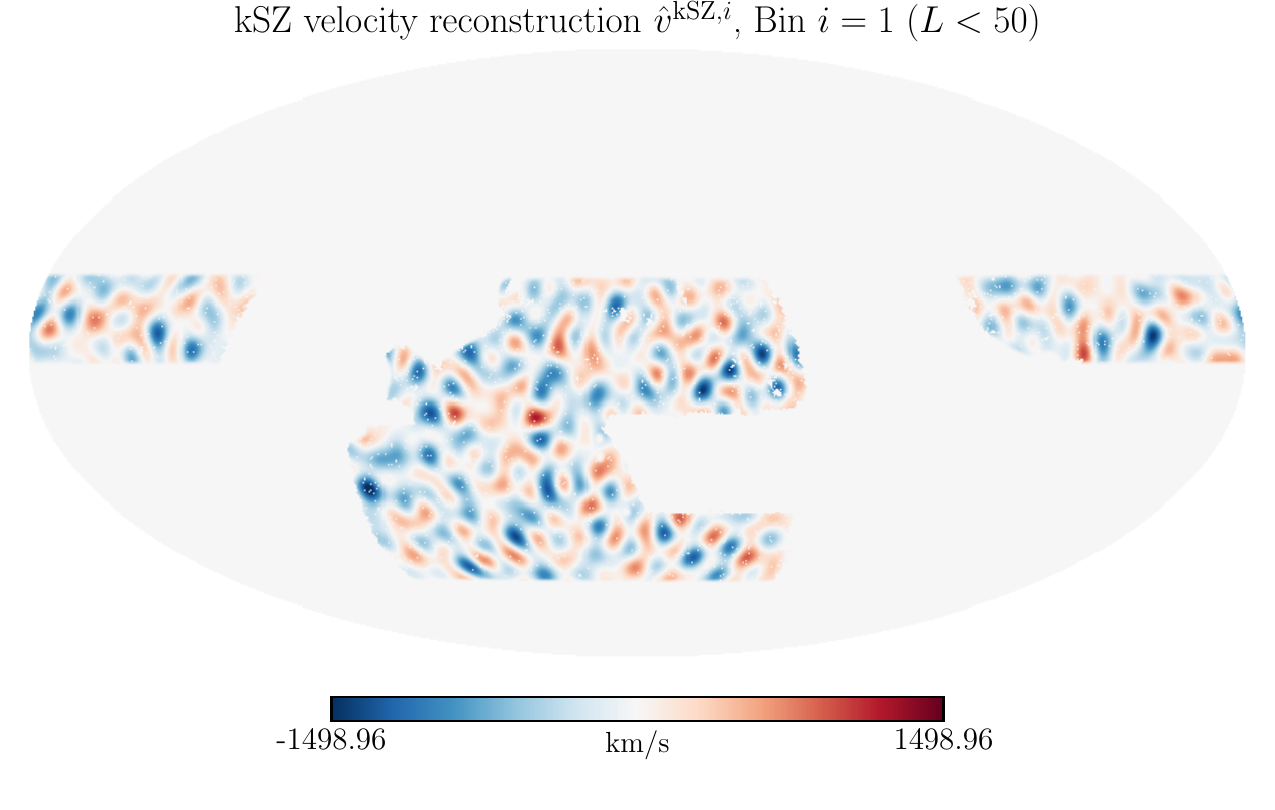}
\includegraphics[width=0.49\textwidth]{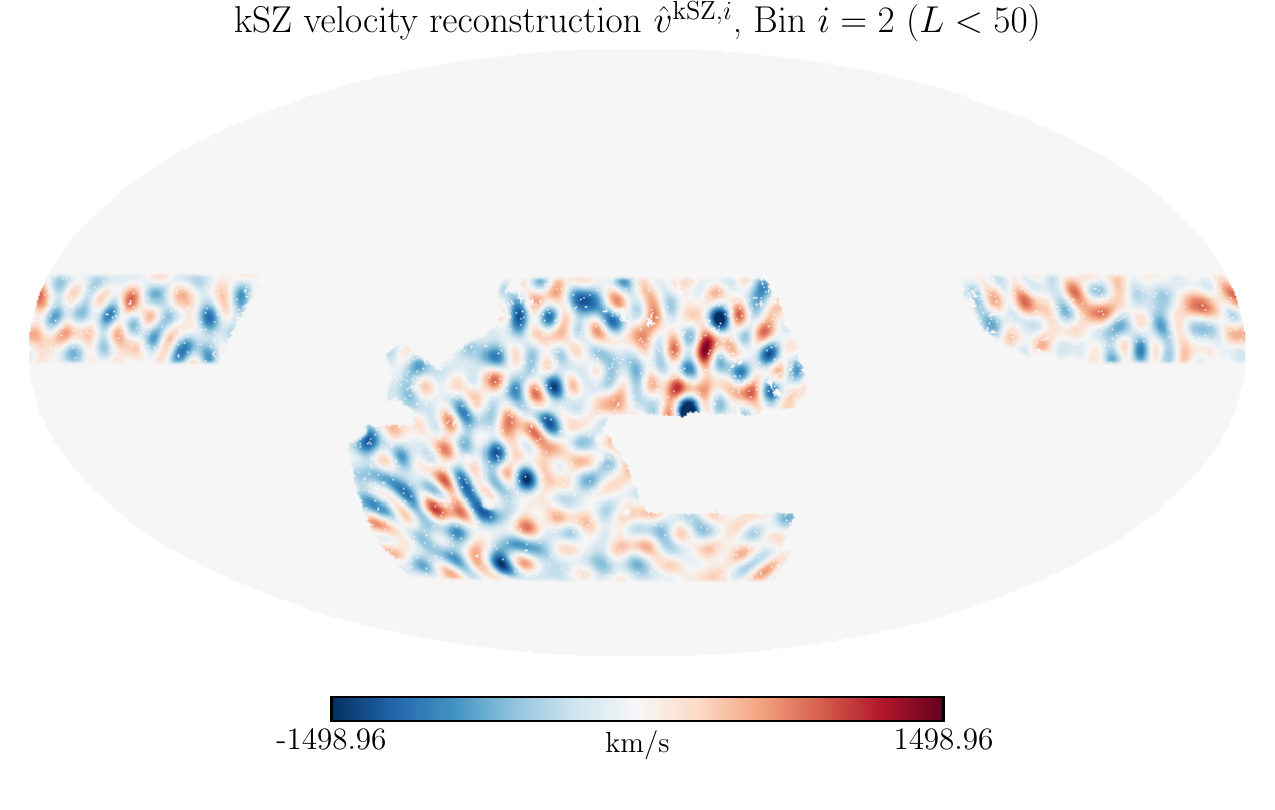}
\includegraphics[width=0.49\textwidth]{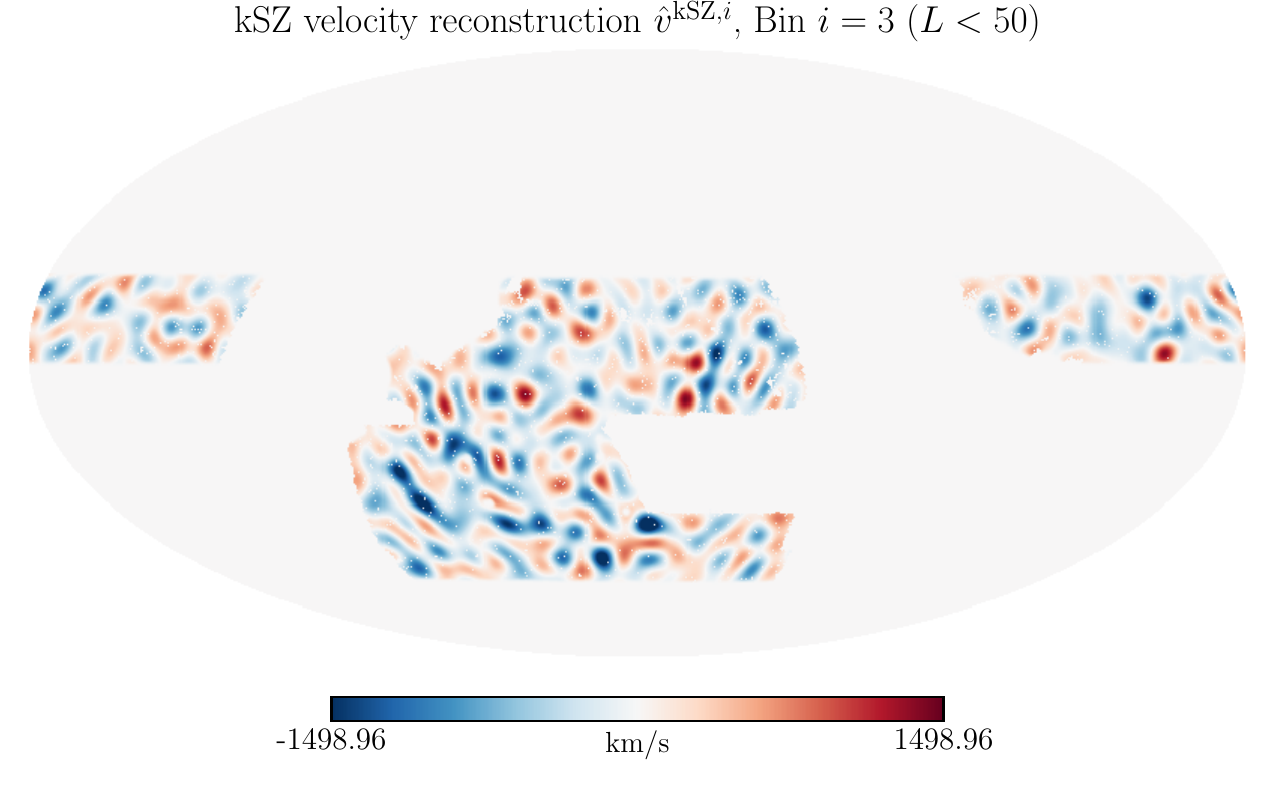}
\includegraphics[width=0.49\textwidth]{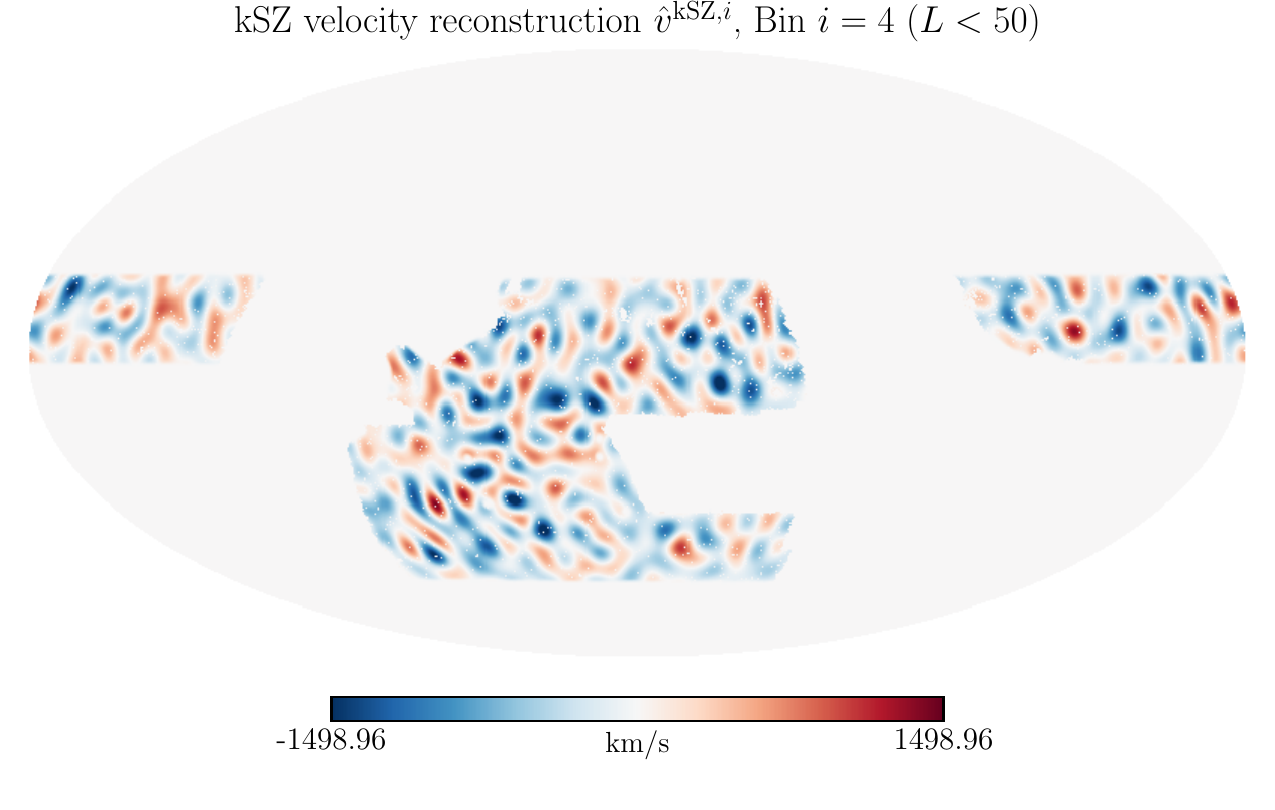}
\caption{The reconstructed velocity, filtered to preserve only $L<50$ information.}\label{fig:velocity_reconstructed_maps_l50}
\end{figure*}

In the main text, we showed the reconstructed velocity  maps filtered to preserve only the $L<20$ information (Figure~\ref{fig:velocity_reconstructed_maps_l20}). In this Appendix, we show analagous plots for $L<10$ and $L<50$ information. These are presented in Figs.~\ref{fig:velocity_reconstructed_maps_l10} and~\ref{fig:velocity_reconstructed_maps_l50} respectively.

% Don't change these lines
\bsp	% typesetting comment
\label{lastpage}
\end{document}